\documentclass[useAMS,usenatbib]{mn2e}
\usepackage[dvips]{graphicx}

 \title[Type IIb Supernova SN 2011dh]{One year of monitoring of the Type IIb 
supernova SN 2011dh}

\author[D. K. Sahu, G. C. Anupama and N. K. Chakradhari ]
{D. K. Sahu$^1$\thanks{E-mail : dks@iiap.res.in (DKS)},
G. C. Anupama$^1$\thanks{E-mail : gca@iiap.res.in (GCA)}, N. K. Chakradhari$^2$\\
1. Indian Institute of Astrophysics, Koramangala, Bangalore 560 034, India\\
2. Pt. Ravi Shankar Shukla University, Raipur, India }

\begin{document}

\date{Accepted .....; Received ......}

\maketitle

\label{firstpage}

\begin{abstract}
Optical $UBVRI$ photometry and low resolution spectroscopy of the type IIb 
supernova SN 2011dh in M51 are presented, covering the first year after the 
explosion. The light curve and spectral evolution are discussed.  The early 
phase light curve evolution of SN 2011dh is very similar to SN 1993J and 
SN 2008ax. In the late phase, however, SN 2011dh declines faster than SN 1993J. 
The late phase decline in the $B$-band is steeper than in the $R$ and $I$
bands, indicating the possibility of dust formation.
With a peak $V$-band absolute magnitude of $M_V = -17.123\pm0.18$ mag, SN 2011dh 
is a marginally faint type IIb event. The reddening corrected colour curves of 
SN 2011dh are found to be redder than other well studied type IIb supernovae. 
 The bolometric light curve indicates $\sim$ 0.09 M$_\odot$ of 
$^{56}$Ni is synthesized during the explosion. The HeI 
lines were detected in the spectra during the rise to maximum. The nebular 
spectra of SN 2011dh show a box shaped emission in the red wing of the [OI] 
6300-6363 \AA\ feature, that is attributed to H$\alpha$ emission from a shock 
excited circumstellar material. The analysis of nebular spectra indicates that 
$\sim 0.2$ M$_\odot$ of oxygen was ejected during the explosion. Further, the 
[CaII]/[OI] line ratio in the nebular phase is $\sim$ 0.7, indicating a 
progenitor with a main sequence mass of 10-15 M$_\odot$.    
\end{abstract}

\begin{keywords}
supernovae: general - supernovae: individual: SN 2011dh - techniques: photometric - 
techniques: spectroscopic
\end{keywords}

\section{Introduction}

Core-collapse supernovae result from the violent death of massive stars with 
initial masses greater than 8 M$_\odot$. Considerable diversity is observed in 
the photometric and spectroscopic properties of these objects, leading to their
classification into various classes. The core collapse supernovae that show the
presence of strong hydrogen lines in their spectra close to maximum light are
designated as type II. The hydrogen deficient core collapse supernovae are
designated type Ib or type Ic based on the presence, or absence of helium lines,
respectively, at light maximum. A further subclassification of type II 
supernovae into type IIP (long plateau in the light curve) and type IIL (linear 
decline after peak brightness) is done based on their light curves.   However, 
there are some transitional events e.g. SN 1987K \citep{filippenko88},  SN 1993J (\citealt{prabhu95},
\citealt{baron95}), SN 2008ax (\citealt{pastorello08}, \citealt{chornock11}), whose 
spectra at the early phase are dominated by H lines, very similar to the type II
supernovae, but in the later phases, their spectra are dominated by He lines, 
similar to the type Ib events. Because of their simliarity with type II 
supernovae in the early phase and later exhibiting properties of type Ib 
supernovae, these transitional objects are classified as type IIb supernovae 
\citep{woosley87}.  The number of supernovae identifed as IIb events is limited,
although recently   these objects  gained greater interest as 
they provide a link between the hydrogen rich type II (type IIP and type IIL) 
supernovae and the hydrogen deficient type Ib and Ic objects. The type Ib, Ic 
and IIb supernovae are collectively known as stripped envelope core-collapse supernovae (CCSNe).

There is a growing consensus that the observed diversity in the properties of 
core-collapse supernovae is due to the state of the progenitor star's hydrogen 
and helium envelopes at the time of explosion (\citealt{filippenko97}, 
\citealt{heger03}, \citealt{gal-yam07}). The subclassification of core-collapse
supernovae can be represented in the form of a sequence - 
IIP$\rightarrow$IIL$\rightarrow$IIb$\rightarrow$Ib$\rightarrow$Ic,  
and can possibly be interpreted 
as a sequence of stripping of the H envelope. The progenitors of type IIP 
supernovae retain most of the H envelope, while the progenitor stars that lose
almost all of the H envelope and retain only a very thin layer of H at the time 
of explosion produce the type IIb events. The progenitors that have lost their 
entire hydrogen envelope produce the type Ib and those that have been stripped 
off hydrogen and most of their helium result in type Ic supernovae. In fact,
detailed modeling of  type IIb supernovae have shown that their observed light 
curves can be reproduced well using helium stars with a very thin hydrogen 
envelope \citep{mazzali09}. 
   
In a few cases, progenitors of nearby core-collapse supernovae have been 
identified in the high resolution, pre-explosion images obtained by the Hubble 
Space Telescope (HST) and other large facilities. The detected progenitors 
of type IIP supernovae indicate that they come from red supergiants (e.g.\ 
\cite{smartt09} and references therein). For the type IIb supernovae, the 
situation is not very clear. Massive stars in close binary systems experiencing 
strong mass transfer \citep{podsiadlowski93}, or very massive single stars with strong stellar winds 
\citep{heger03} are considered as potential progenitors of type IIb events. The 
progenitor system of SN 1993J was identified as a K0Ia star \citep{filippenko93}
in a binary system, with an early B-supergiant companion \citep{maund09}.
Similarly, an interacting  binary system, with a slightly later companion star 
(late B through late F), was suggested as the progenitor of SN 2001ig 
\citep{ryder06}. A  source, coinciding with the position of SN 2008ax was 
identified in pre-explosion HST images (\citealt{li08}, \citealt{crockett08}).
However, its interpretation as the progenitor is ambiguous. \cite{crockett08} 
have explored the possibility of (i) the progenitor being a single massive star 
that lost most of its hydrogen envelope through radiatively driven mass loss 
processes,  before exploding as a helium-rich Wolf -Rayet star, or (ii) a stripped
star in an interacting binary in a low mass cluster. The [Ca\,II]/[O\,I] ratio in 
nebular spectra of SN 2008ax, however, is more consistent with the low mass
binary scenario \citep{taubenberger11}.

A new supernova was discovered by A. Riou on 2011 June 1.893 in the nearby 
galaxy M51 \citep{griga11}. Images of the galaxy obtained on May 31.893 also 
showed the supernova, but, no object was visible at the position of the 
supernova in the images of May 30. The Palomar Transient Factory (PTF) 
independently discovered the supernova on 2011 June 1.19 \citep{silverman11},
while the PTF observations of May 31.275 do not show the supernova. These 
observations reduce the uncertainty on the date of explosion to better than 
0.6 days \citep{arcavi11}. Based on the earliest spectrum of 2011 June 3, the 
supernova was classified as a young type II supernova (\citealt{silverman11}, 
\citealt{yamanaka11}).  \cite{arcavi11a} noticed the similarity between the 
spectrum of SN 2011dh and the type IIb events SN 1993J and SN 2008ax, and 
suggested SN 2011dh was possibly a type IIb event.  Further observations on
June 12 and 16 by \cite{marion11} in the infra-red showed the presence of
He\,I features consistent with its classification as a type IIb event.  

SN 2011dh has been followed extensively in a wide wavelength range, from the
radio to the X-rays. It was detected at 86 GHz just three days after its 
discovery, by the Combined Array for Research in Millimeter-wave Astronomy 
(CARMA) \citep{horesh11} and has been followed with VLBI (\citealt{marti11}, 
\citealt{bietenholz12}) and EVLA over the first 100 days of its evolution 
\citep{krauss12}. SN 2011dh was also detected in the X-rays by the {\it 
Swift}/X-ray Telescope (XRT) $\sim$ 3 days after the explosion, and has
subsequently been followed with both {\it Swift} and {\it Chandra}. A 
multi-wavelength study of SN 2011dh spanning the radio, milimeter, X-ray and 
gamma-ray bands during the first few seconds to weeks following the explosion 
is presented by \cite{soderberg12}. \cite{maund11} have presented the optical 
photometric and spectroscopic data of SN 2011dh during the first 50 days after 
explosion, while photometric data covering the first $\sim$ 300 days after 
explosion, with a preliminary light curve modelling, have been presented by 
\cite{tsvetkov12}. 

The reddening in the Milky Way in the direction of M51 is $E(B-V)_{gal}= 0.035$ mag 
\citep{schlegel98}. High resolution spectroscopy of SN 2011dh did not show 
any narrow lines due to interstellar medium in the host galaxy, except the 
Na\,ID doublet due to the interstellar medium in the Milky Way 
(\citealt{vinko12}, \citealt{arcavi11}, \citealt{ritchey12}), indicating low
extinction within the host galaxy. Hence, $E(B-V) = 0.035$ mag is adopted as 
 total reddening. 

A search for the progenitor star in the archival pre-explosion HST images 
obtained with the Advanced Camera for Survey led to the detection of a luminous 
star at the position of the supernova (\citealt{vandyk11}, \citealt{maund11}, 
\citealt{murphy11}), but its nature and association with the supernova remains  
controversial. The SED of this candidate progenitor star was found to be 
consistent with an F8 supergiant, but with a higher luminosity and a more 
extended radius than a normal supergiant (\citealt{vandyk11}, \citealt{maund11},
\citealt{vinko12}). \cite{maund11} and \cite{vandyk11} estimate the initial mass
of this proposed progenitor to be $M_{ZAMS} \sim 15-20$ M$_\odot$. The observed 
early optical, radio and X-ray observations of SN 2011dh (\citealt{arcavi11}, 
\citealt{soderberg12}, \citealt{vandyk11}) and the re-analyses of the HST images
with an improved distance to M51 \citep{vinko12} all point toward a compact 
progenitor, while hydrodynamical modelling suggests that a large progenitor star
with radius $\sim$ 200 $R\odot$ is required to reproduce the early light curve 
\citep{bresten12}. The association of the yellow supergiant with SN 2011dh 
needs further exploration. \cite{benvenuto13} explored the possibility of the
existence of such a star in a close binary system by performing binary stellar
evolution calculations and arrived at the conclusion that a close binary system
of solar composition stars with masses of 16 M$_\odot + 10$ M$_\odot$ could  lead
to the evolution of the donor star into a yellow supergiant consistent with the 
candidate progenitor detected in the HST images.

In this paper we present a detailed analysis of optical photometry in 
$U$, $B$, $V$, $R$, and $I$ bands and medium resolution spectroscopy of 
SN 2011dh during the first one year after the explosion, obtained with the 2m 
Himalayan Chandra Telescope of the Indian Astronomical Observatory, Hanle, 
India. 

\section{Observation and Data reduction}
\subsection{Photometry}

Photometric monitoring of SN 2011dh was carried out with the 2m Himalayan Chandra 
Telescope (HCT) of the Indian Astronomical Observatory (IAO), Hanle, India. The 
imaging observations were made in Bessell's $U$,  $B$, $V$, $R$ and $I$ filters 
available with the Himalaya Faint Object Spectrograph Camera (HFOSC). HFOSC is 
equipped with a 2k$\times$4k pixels, SITe CCD chip. The central 2k$\times$2k 
pixels of the chip was used for imaging observations. Plate scale of the 
telescope-detector system is 0.296$''$/pixel and the central 2k$\times$2k 
pixels of the chip covers 10$'$$\times$10$'$ region of the sky.
 
Photometric monitoring of SN 2011dh started on 2011 June 03 (JD 245\,5716.20) 
and continued till 2012 May 20 (JD 245\,6068.36), with a break in between,
during 2011 September 06  to 2011 November 19, when the object was in Solar 
conjunction. The object was monitored in $B$, $V$, $R$ and $I$ bands during the 
entire period, whereas the observations in $U$-band could be obtained only upto
November 19, 2011. Photometric standard fields from Landolt's catalogue 
\citep{landolt92} observed on 2011 June 3, under photometric conditions were 
used for calibration. Standard fields observed during 2005-2006 as a part of an 
extensive monitoring campaign of SN 2005cs, also hosted in M51, were also used,
together with the observations of 2011 June 3, for photometric calibration of a 
sequence of secondary standards in the supernova field. 

\begin{figure}
\resizebox{\hsize}{!}{\includegraphics{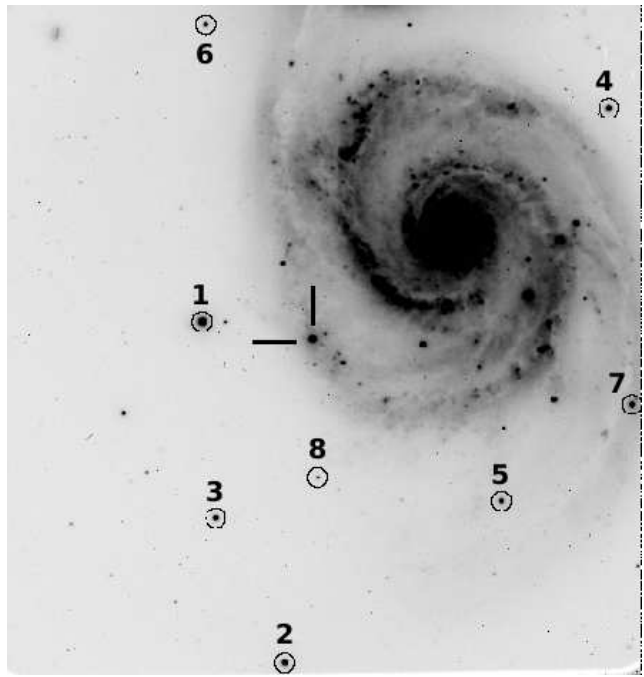}}
\caption[]{Identification chart for SN 2011dh. The stars used as local standards are
marked with numbers 1-8. North is up and east to the left. The field of view
is $10^\prime \times 10^\prime$.}
\label{fig_std}
\end{figure}

All the imaging data were pre-processed in the standard way using various tasks 
available within IRAF\footnote{IRAF is distributed by the National Optical 
Astronomy Observatories, which are operated by the Association of Universities 
for Research in Astronomy, Inc., under cooperative agrement with the National 
Science Foundation}. Instrumental magnitudes of the standard stars were obtained
using aperture photometry, with an optimal aperture, which is usually 3-4 times 
the full width half maximum (FWHM) of the stellar profile, determined using the 
aperture growth curve. Aperture correction between the optimal aperture and an 
aperture close to the FWHM of the stellar profile that had the maximum 
signal-to-noise ratio was determined using the bright stars in the field and 
then applied to the fainter ones. Correction for atmospheric extinction was made
using the average extinction values for the site \citep{stalin08}, and the 
average colour terms for the system were used to determine the photometric zero 
points on individual nights.  These were then used to calibrate a sequence 
of local standards in the supernova field (marked in Figure \ref{fig_std})
observed on the same nights as the standard fields. The $U$, $B$, $V$, $R$ and 
$I$ magnitudes of the secondary standards, averaged over four nights are listed 
in Table \ref{tab_std}.  The errors reported with the magnitudes are the 
standard deviation of the standard magnitudes obtained on the four nights.  The 
magnitudes of the secondary standards obtained using the two separate observing 
runs, during 2005-2006 and 2011, are in good agreement. Further, we have four
secondary standards in common with \cite{pastorello09}. The average of the 
absolute value of difference between our magnitudes and those reported by 
\cite{pastorello09} are 0.058, 0.023, 0.030, 0.015 and 0.059 in $U$, $B$, $V$, 
$R$, $I$ bands, respectively. 

\begin{table*}
\caption{Magnitudes for the sequence of secondary standard stars in
the field of SN 2011dh.}
\begin{tabular}{lccccc}
\hline\hline
ID & U  & B & V &  R & I \\
\hline\hline     
1 & 14.689$\pm$0.013 & 14.339$\pm$0.015& 13.640$\pm$ 0.023&   13.223$\pm$ 0.024&  12.899$\pm$ 0.048\\	
2 & 15.496$\pm$0.023 & 15.473$\pm$0.028& 14.851$\pm$ 0.027&   14.498$\pm$ 0.007&  14.100$\pm$ 0.006\\
3 & 17.331$\pm$0.032 & 16.561$\pm$0.022& 15.644$\pm$ 0.028&   15.095$\pm$ 0.029&  14.636$\pm$ 0.014\\ 
4 & 16.744$\pm$0.047 & 16.216$\pm$0.068& 15.295$\pm$ 0.090&   14.733$\pm$ 0.080&  14.153$\pm$ 0.023\\
5 & 16.249$\pm$0.020 & 16.221$\pm$0.026& 15.605$\pm$ 0.030&   15.233$\pm$ 0.029&  14.866$\pm$ 0.019\\
6 & 17.161$\pm$0.042 & 16.974$\pm$0.036& 16.256$\pm$ 0.003&   15.805$\pm$ 0.031&  15.371$\pm$ 0.016\\
7 & 17.753$\pm$0.037 & 16.586$\pm$0.025& 15.090$\pm$ 0.027&   14.103$\pm$ 0.007&  13.058$\pm$ 0.040\\
8 & 20.589$\pm$0.081 & 19.429$\pm$0.050& 17.926$\pm$ 0.037&   16.799$\pm$ 0.023&  15.540$\pm$ 0.019\\
\hline
\end{tabular}
\label{tab_std}
\end{table*}

The supernova is located at the edge of outer spiral arm, 138$''$ east and 92$''$ south 
of the nucleus of M51. During the brighter phase, profile fitting 
photometry of the supernova is not contaminated much by the underlying galaxy 
background. However, as the supernova dimmed, profile fitting photometry 
overestimated the supernova flux, and the photometric precision also degraded.
Hence, host galaxy templates in different bands, selected from amongst the
best frames of SN 2005cs observations, were used for a more accurate subtraction
of the host galaxy background. The magnitudes of the secondary standards in the 
supernova frames and those of supernova in the templated-subtracted frames were
obtained using aperture photometry. The night-to-night zero-points were 
determined using the stars identified as local standards in the supernova field 
and the supernova magnitudes were calibrated differentially with respect to the 
local standards. The estimated supernova magnitudes in $U$, $B$, $V$, $R$ and 
$I$ bands are listed in Table \ref{tab_snmag}. The errors on the magnitudes were
estimated by adding in quadrature the errors associated with nightly 
photometric zero points, and the fitting errors as computed by IRAF.  

\subsection{Spectroscopy}

Medium resolution optical spectra of SN 2011dh were obtained on 20 epochs. The 
spectra were obtained using grism Gr\#7 (wavelength range $3500 - 7800$ \AA) 
and Gr\#8 (wavelength range $5200 - 9250$ \AA) available with the HFOSC 
instrument. Our spectroscopic monitoring, which started on JD 245\,5716 and 
continued till JD 245\,6072, presents the spectral evolution from $\sim$ 3 days 
to $\sim$ 1 year after the explosion. A journal of spectroscopic observation is
provided in Table \ref{tab_spec}.  Spectrophotometric standards Feige 34, 
Feige 110, Hz 44, Wolf 1346 were observed for determining the instrumental 
response correction. 

All spectra were reduced  in the standard manner using various tasks available 
in IRAF. The two-dimensional spectral frames were bias corrected and 
flat-fielded, and the one dimensional 
spectra were extracted using the optimal extraction method \citep{horne86}. 
Arc lamp spectra were used for applying wavelength calibration. The bright night
sky emission lines were used to cross check the wavelength calibration, and 
whenever required, small shifts were applied to the observed spectra. 
Instrumental response was corrected using the spectra of spectrophotometric 
standard stars observed during the same night. On some nights, when 
spectrophotometric standard stars could not be observed, spectra of standard
stars observed on nearby nights were used for the correction. The flux
calibrated spectra in the blue and red regions were combined after proper 
scaling to get the final spectrum on a relative flux scale. The spectra were 
then brought to an absolute flux scale using zero points determined from 
broad-band $UBVRI$ magnitudes. The supernova spectra were then corrected for 
the host galaxy redshift z =0.002 and dereddened for a total  reddening of 
$E(B-V)=0.035$ mag. The telluric lines have not been removed from the spectra due
to the absence of a template hot star spectrum with a good signal-to-noise
ratio.
  
\section{Results}
\subsection{Photometric results}
\subsubsection{Light curves}

\begin{figure}
\resizebox{\hsize}{!}{\includegraphics{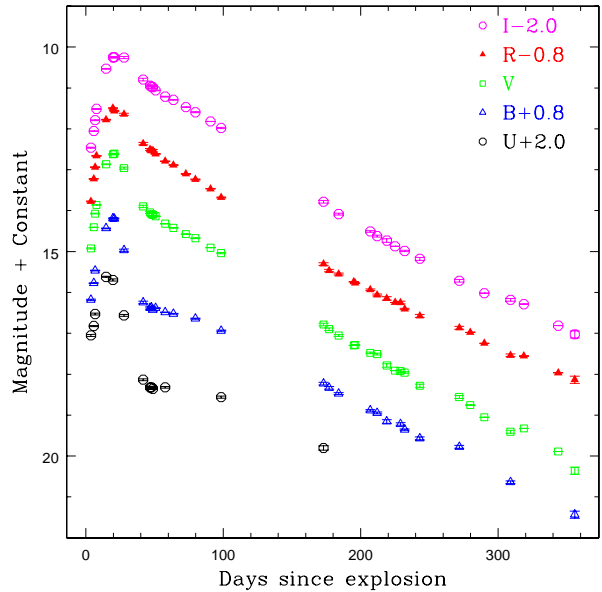}}
\caption[]{$UBVRI$ light curves of SN 2011dh. The light curves have been shifted by the 
amount indicated in the legend. \\
(A colour version of this figure is available in the online journal)}
\label{fig_light}
\end{figure}

The light curves of SN 2011dh in $U$,  $B$, $V$, $R$ and $I$ bands are presented
in Figure \ref{fig_light}. The observed supernova magnitudes were used to 
obtain photometric parameters of SN 2011dh in different bands, and are listed 
in Table \ref{tab_parameter}. The maximum in $B$-band occurred on JD 245\,5732.60$\pm$0.35, 
at an apparent magnitude of $13.388\pm 0.022$. The maximum in $U$-band occurred 
$\sim 4$ days before the $B$-band maximum, whereas the maximum in the $V$, $R$ 
and $I$ bands occurred $\sim$ 1, 1.5 and 3.3 days, respectively, after $B$ 
maximum.  The date of maximum and peak magnitudes in different bands are
reported by Tsvetkov et al (2012). Except for the $U$-band, where our estimate 
of JD maximum differs by $\sim1.5$ days, our estimates of the date of maximum 
and peak observed magnitudes are consistent with those of \cite{tsvetkov12}. 
A reddening of $E(B-V)=0.035$ mag, and a distance of $8.4\pm0.7$ Mpc 
\citep{vinko12} are used in obtaining the peak absolute magnitudes of SN 
2011dh. The errors in the absolute magnitudes have been arrived at by using 
uncertainties in the peak magnitude and the distance modulus of the host galaxy.
The rise time to maximum in different bands is also listed in Table 
\ref{tab_parameter}. The rise time of SN 2011dh in different bands is very 
similar to that of SN 2008ax (\citealt{pastorello08}, \citealt{taubenberger11}).

\begin{table*}
\caption{Photometric observations of SN 2011dh}
\begin{tabular}{lcccccccc}
\hline\hline
 Date & J.D. & Phase\rlap{*} & U & B & V & R & I\\
     &   & (days)    &   &   &   &   &\\
\hline\hline

03/06/2011& 245\,5716.20&3.20&  15.051$\pm$0.024 &  15.379$\pm$0.014 &  14.924$\pm$0.016 &  14.576$\pm$0.019 &  14.461$\pm$0.014\\    
05/06/2011& 245\,5718.22&5.22&  14.821$\pm$0.013 &  14.972$\pm$0.004 &  14.407$\pm$0.025 &  14.025$\pm$0.017 &  14.054$\pm$0.023\\    
06/06/2011& 245\,5719.22&6.22&  14.528$\pm$0.020 &  14.666$\pm$0.010 &  14.074$\pm$0.018 &  13.738$\pm$0.020 &  13.787$\pm$0.019\\    
07/06/2011& 245\,5720.22&7.22&                   &                   &  13.862$\pm$0.014 &  13.459$\pm$0.010 &  13.513$\pm$0.025\\    
14/06/2011& 245\,5727.32&14.32&  13.620$\pm$0.013 &  13.632$\pm$0.010 &  12.867$\pm$0.014 &  12.577$\pm$0.026 &  12.530$\pm$0.017\\   
19/06/2011& 245\,5732.28&19.28&  13.694$\pm$0.026 &  13.389$\pm$0.012 &  12.624$\pm$0.020 &  12.294$\pm$0.009 &  12.253$\pm$0.019\\   
20/06/2011& 245\,5733.23&20.33&                   &  13.394$\pm$0.013 &  12.608$\pm$0.017 &  12.363$\pm$0.034 &  12.250$\pm$0.192\\   
27/06/2011& 245\,5740.28&27.28&  14.561$\pm$0.022 &  14.167$\pm$0.027 &  12.960$\pm$0.025 &  12.445$\pm$0.014 &  12.256$\pm$0.032\\   
11/07/2011& 245\,5754.21&41.21&  16.133$\pm$0.022 &  15.442$\pm$0.031 &  13.897$\pm$0.037 &  13.161$\pm$0.011 &  12.795$\pm$0.022\\   
16/07/2011& 245\,5759.23&46.23&  16.326$\pm$0.031 &  15.582$\pm$0.024 &  14.044$\pm$0.016 &  13.303$\pm$0.029 &  12.947$\pm$0.014\\   
17/07/2011& 245\,5760.21&47.21&  16.321$\pm$0.016 &  15.570$\pm$0.035 &  14.089$\pm$0.022 &  13.336$\pm$0.036 &  12.971$\pm$0.018\\   
18/07/2011& 245\,5761.23&48.23&  16.354$\pm$0.014 &  15.624$\pm$0.029 &  14.101$\pm$0.016 &  13.333$\pm$0.041 &  12.984$\pm$0.014\\   
20/07/2011& 245\,5763.18&50.18&                   &  15.583$\pm$0.009 &  14.135$\pm$0.013 &  13.413$\pm$0.024 &  13.062$\pm$0.013\\   
27/07/2011& 245\,5770.18&57.18&  16.320$\pm$0.018 &  15.685$\pm$0.007 &  14.319$\pm$0.024 &  13.590$\pm$0.027 &  13.213$\pm$0.018\\   
02/08/2011& 245\,5776.20&63.20&                   &  15.723$\pm$0.010 &  14.421$\pm$0.019 &  13.685$\pm$0.021 &  13.292$\pm$0.019\\   
11/08/2011& 245\,5785.13&72.13&                   &                   &  14.574$\pm$0.032 &  13.900$\pm$0.019 &  13.468$\pm$0.012\\   
18/08/2011& 245\,5792.13&79.13&                   &  15.842$\pm$0.012 &  14.673$\pm$0.027 &  14.039$\pm$0.016 &  13.594$\pm$0.013\\   
29/08/2011& 245\,5803.14&90.14&                   &                   &  14.907$\pm$0.018 &  14.268$\pm$0.017 &  13.820$\pm$0.018\\   
06/09/2011& 245\,5811.09&98.09&  16.560$\pm$0.024 &  16.135$\pm$0.014 &  15.038$\pm$0.030 &  14.478$\pm$0.007 &  13.979$\pm$0.018\\   
19/11/2011& 245\,5885.50&172.50&  17.804$\pm$0.055 &  17.427$\pm$0.033 &  16.784$\pm$0.032 &  16.105$\pm$0.021 &  15.784$\pm$0.024\\   
24/11/2011& 245\,5889.51&176.51&                   &  17.534$\pm$0.032 &  16.898$\pm$0.027 &  16.265$\pm$0.044 &                  \\   
30/11/2011& 245\,5896.48&183.48&                   &  17.673$\pm$0.023 &  17.053$\pm$0.026 &  16.348$\pm$0.017 &  16.084$\pm$0.058\\   
12/12/2011& 245\,5907.52&194.52&                   &                   &  17.297$\pm$0.024 &  16.540$\pm$0.020 &                  \\    
13/12/2011& 245\,5908.53&195.53&                   &                   &  17.284$\pm$0.035 &  16.572$\pm$0.021 &                  \\   
23/12/2011& 245\,5919.43&206.43&                   &  18.085$\pm$0.020 &  17.478$\pm$0.021 &  16.734$\pm$0.024 &  16.510$\pm$0.057\\   
28/12/2011& 245\,5924.47&211.47&                   &  18.147$\pm$0.022 &  17.506$\pm$0.039 &  16.862$\pm$0.018 &  16.624$\pm$0.035\\   
04/01/2012& 245\,5931.40&218.40&                   &  18.357$\pm$0.044 &  17.786$\pm$0.039 &  16.945$\pm$0.033 &  16.731$\pm$0.034\\   
10/01/2012& 245\,5937.48&224.48&                   &                   &  17.913$\pm$0.041 &  17.043$\pm$0.023 &  16.871$\pm$0.025\\   
14/01/2012& 245\,5941.49&228.49&                   &  18.416$\pm$0.031 &  17.921$\pm$0.021 &  17.048$\pm$0.032 &                  \\   
18/01/2012& 245\,5944.51&231.51&                   &  18.553$\pm$0.018 &  17.959$\pm$0.028 &  17.204$\pm$0.019 &  16.986$\pm$0.029\\   
28/01/2012& 245\,5955.47&242.47&                   &  18.766$\pm$0.036 &  18.281$\pm$0.015 &  17.375$\pm$0.025 &  17.178$\pm$0.023\\   
26/02/2012& 245\,5984.34&271.34&                   &  18.974$\pm$0.034 &  18.557$\pm$0.040 &  17.666$\pm$0.019 &  17.714$\pm$0.080\\   
05/03/2012& 245\,5992.28&279.28&                   &                   &  18.753$\pm$0.052 &  17.775$\pm$0.029 &                  \\   
15/03/2012& 245\,6002.47&289.47&                   &                   &  19.052$\pm$0.028 &  18.044$\pm$0.022 &  18.018$\pm$0.033\\   
03/04/2012& 245\,6021.41&308.41&                   &  19.839$\pm$0.033 &  19.406$\pm$0.027 &  18.337$\pm$0.048 &  18.181$\pm$0.060\\   
13/04/2012& 245\,6031.31&318.31&                   &                   &  19.326$\pm$0.058 &  18.354$\pm$0.034 &  18.287$\pm$0.040\\   
08/05/2012& 245\,6056.29&343.29&                   &                   &  19.890$\pm$0.047 &  18.765$\pm$0.038 &  18.812$\pm$0.078\\   
20/05/2012& 245\,6068.36&355.36&                   &  20.632$\pm$0.088 &  20.360$\pm$0.044 &  18.933$\pm$0.041 &  19.025$\pm$0.070\\   
\hline
\multicolumn{8}{l}{\rlap{*}\ \  Observed phase with respect to the date of explosion  (JD 245\,5713.0).}
\end{tabular}			    
\label{tab_snmag}	    
\end{table*}

\begin{table*}
\caption{Log of spectroscopic observations of SN 2011dh.}
\begin{tabular}{lccc}
\hline\hline
Date & J.D. & Phase\rlap{*} & Range \\
     & 2450000+ & days & \AA\  \\
\hline\hline
03/06/2011&      245\,5716.29 &   3.29   &          3500-7000;5200-9100\\
06/06/2011&      245\,5719.18 &   6.18   &          3500-7000;5200-9100\\
19/06/2011&      245\,5732.29 &   19.29  &          3500-7000;5200-9100\\
20/06/2011&      245\,5733.24 &   20.24  &          3500-7000;5200-9100\\
27/06/2011&      245\,5740.29 &   27.29  &          3500-7000;5200-9100\\
11/07/2011&      245\,5754.22 &   41.22  &          3500-7000;5200-9100\\
17/07/2011&      245\,5760.23 &   47.23  &          3500-7000;5200-9100\\
26/07/2011&      245\,5769.11 &   56.11  &          3500-7000;5200-9100\\
02/08/2011&      245\,5776.21 &   63.21  &          3500-7000;5200-9100\\
07/09/2011&      245\,5812.08 &   99.08  &          3500-7000;5200-9100\\
20/11/2011&      245\,5886.48 &   173.48 &          3500-7000;5200-9100\\ 
12/12/2011&      245\,5908.48 &   195.48 &          3500-7000;5200-9100\\
28/12/2011&      245\,5923.51 &   210.51 &          3500-7000;5200-9100\\
07/01/2012&      245\,5934.31 &   221.31 &          3500-7000;5200-9100\\
10/01/2012&      245\,5937.34 &   224.34 &          3500-7000;5200-9100\\
28/01/2012&      245\,5955.44 &   242.44 &          3500-7000\\
26/02/2012&      245\,5984.39 &   271.39 &          3500-7000;5200-9100\\
05/03/2012&      245\,5992.34 &   279.34 &          3500-7000;5200-9100\\
08/05/2012&      245\,6056.37 &   343.37 &          3500-7000;5200-9100\\
24/05/2012&      245\,6072.18 &   359.18 &          3500-7000;5200-9100\\
\hline
\multicolumn{4}{l}{\rlap{*}\ \  Observed phase with respect to the date of explosion  (JD 245\,5713.0).}
\end{tabular}
\label{tab_spec}
\end{table*}

\begin{table*}
\caption{Photometric parameters}
\begin{tabular}{lcccc}
\hline\hline
Band & JD (Max)  & Peak obs. mag & Peak abs. mag &  Rise time (days)  \\
\hline\hline     
$U$ & 245\,5728.82$\pm$0.43  &     13.605$\pm$0.037   &    -16.186$\pm$0.18 &  15.8$\pm$0.5\\
$B$ & 245\,5732.60$\pm$0.35  &     13.388$\pm$0.022   &    -16.378$\pm$0.18 &  19.6$\pm$0.5\\                 
$V$ & 245\,5733.57$\pm$0.28  &     12.607$\pm$0.016   &    -17.123$\pm$0.18 &  20.6$\pm$0.5\\
$R$ & 245\,5734.11$\pm$0.20  &     12.270$\pm$0.032   &    -17.433$\pm$0.18 &  21.3$\pm$0.5\\
$I$ & 245\,5735.92$\pm$0.31  &     12.196$\pm$0.035   &    -17.477$\pm$0.18 &  22.9$\pm$0.5\\
\hline
\end{tabular}
\label{tab_parameter}
\end{table*}

\begin{figure}
\resizebox{\hsize}{!}{\includegraphics{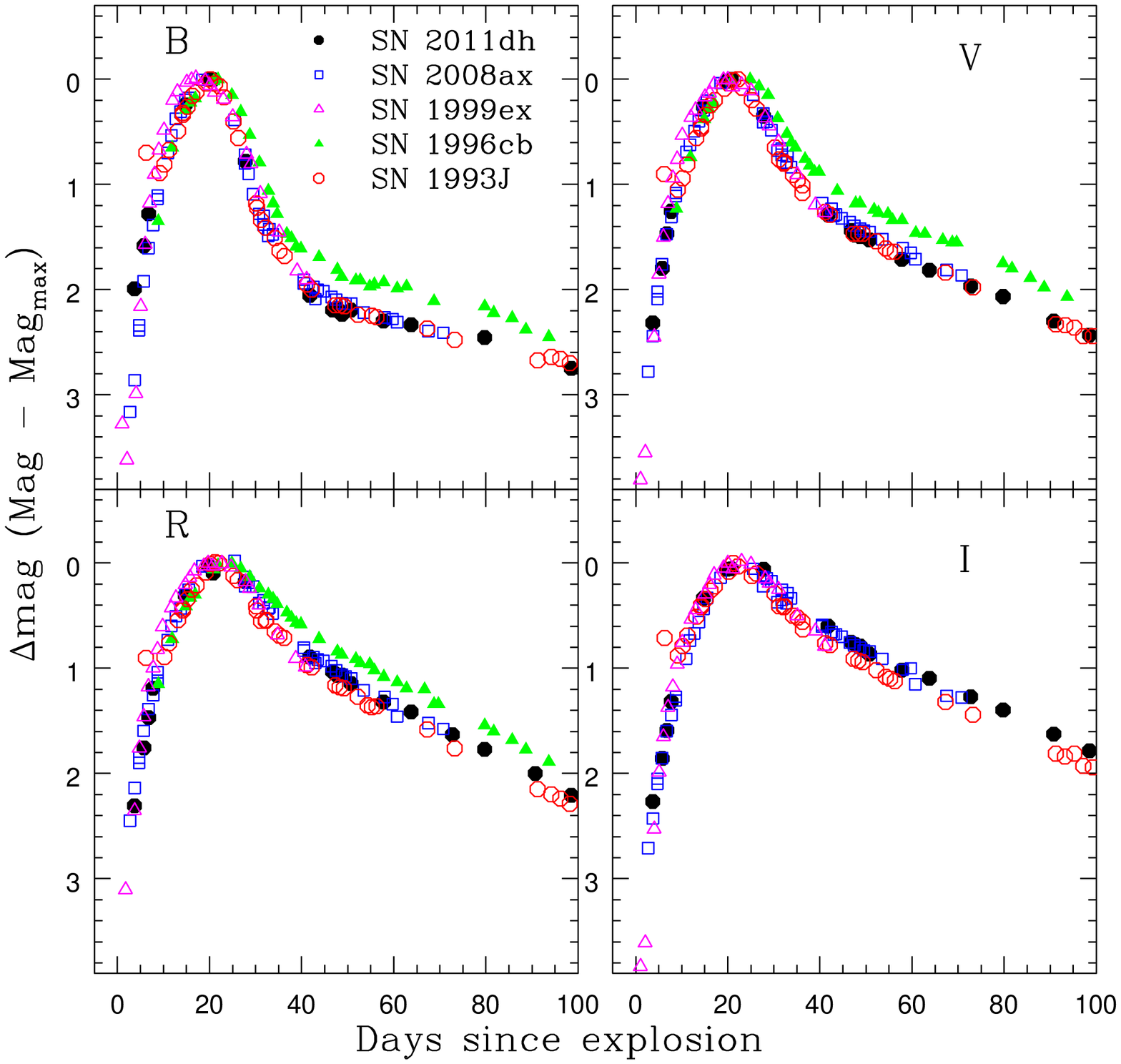}}
\caption[]{Comparison of $UBVRI$ light curves of SN 2011dh with those of 
SN 2008ax, SN 1999ex, SN 1996cb and SN 1993J, during the early phase. The light curves have 
been normalized as described in the text.\\
(A colour version of this figure is available in the online journal.) }
\label{fig_lcomp1}
\end{figure}

\begin{figure}
\resizebox{\hsize}{!}{\includegraphics{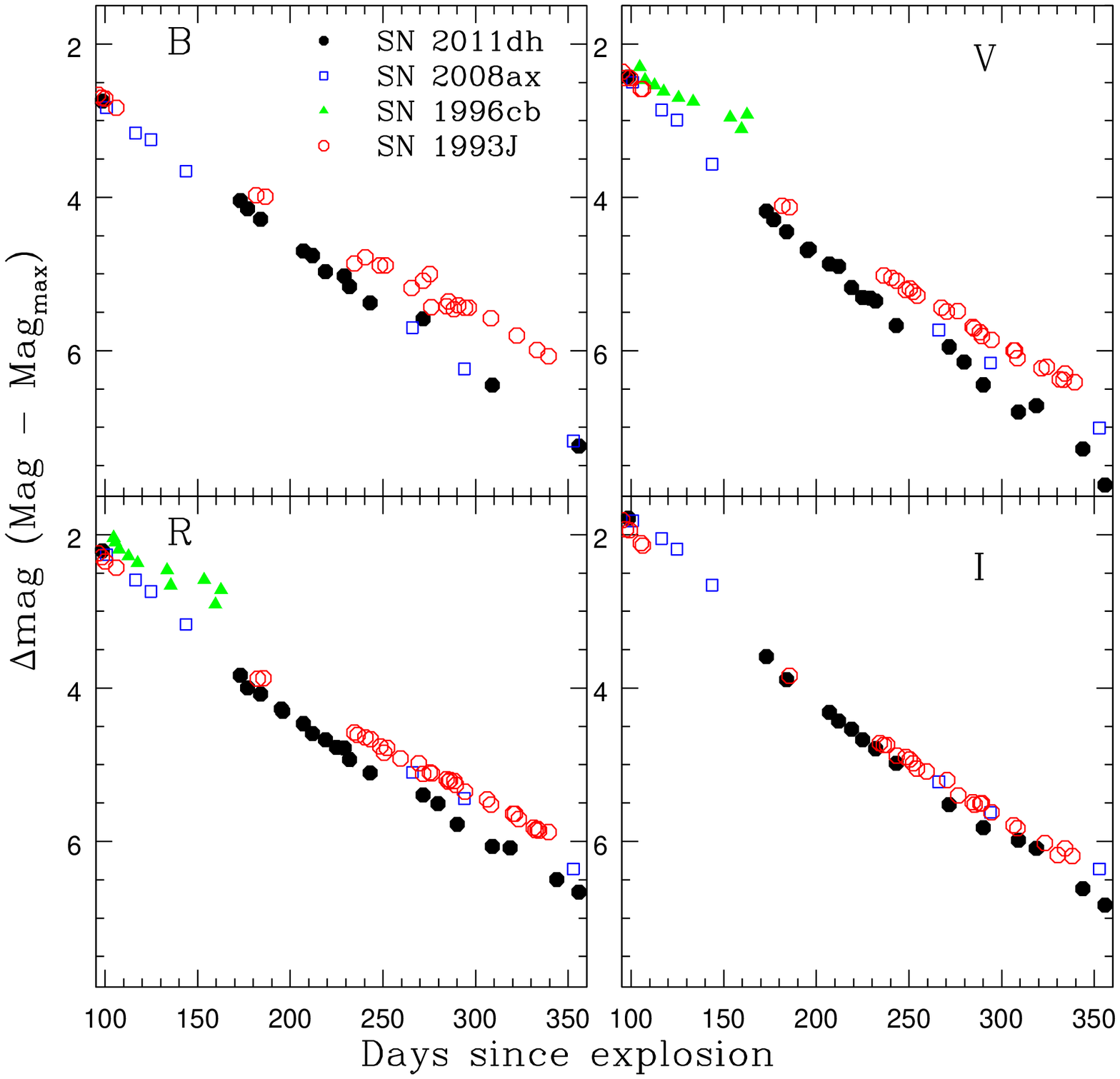}}
\caption[]{Comparison of $UBVRI$ light curves of SN 2011dh with those of 
SN 200ax, SN 1996cb and SN 1993J, during the late phase. The light curves have 
been normalized as described in the text. \\
(A colour version of this figure is available in the online journal.)}
\label{fig_lcomp2}
\end{figure}

Light curves of SN 2011dh are compared with those of type IIb supernovae 
2008ax (\citealt{pastorello08}, \citealt{taubenberger11}), 1993J 
(\citealt{lewis94}, \citealt{baron95}),  1996cb \citep{qui99} and the  type Ib
SN 1999ex \citep{stritzinger02} in Figures \ref{fig_lcomp1} and \ref{fig_lcomp2}. Figure 
\ref{fig_lcomp1} shows the evolution of the light curves during the first 100 
days after explosion and Figure \ref{fig_lcomp2} shows the light curve evolution
beyond 100 days after explosion. The observed magnitudes of the supernovae have 
been normalized to their respective peak magnitudes and shifted in time to 
the epoch of maximum brightness in $B$-band. The light curves of SN 2011dh in 
the $B$ and $V$ bands are very similar to those of SNe 2008ax, 1999ex and 1993J.
The decline in magnitude within 15 days from the date of maximum 
($\Delta m_{15}$), for the $B$, $V$, $R$ and $I$ bands is estimated to be 
$\Delta m_{15}(B) = 1.75 \pm0.18$, $\Delta m_{15}(V) = 0.98\pm0.04$, 
$\Delta m_{15}(R)=0.64\pm0.03$  and $\Delta m_{15}(I)=0.47\pm0.01$. 
\cite{tsvetkov12} estimate $\Delta m_{15}(B)= 1.64$ mag, which is 
consistent with our estimate. These values are marginally
larger than the $\Delta m_{15}$ values of SN 2008ax reported by 
\cite{taubenberger11}, but similar to the $\Delta m_{15}$ values of SN 1993J 
in the $B$ and $V$ bands. 

 The slope of the light curve in the $B$, $V$, $R$ and $I$ bands during
$\sim 60-100$ is estimated to be 1.09$\pm$0.15, 1.76$\pm$0.04, 2.16$\pm$0.05 and 
1.90$\pm$0.05 mag (100 day)$^{-1}$, respectively.
A change in the slope is noticed in all the bands during days $\sim 170-360$.
The $B$ and $V$ light curves show a steepening, with the slope being 
1.71$\pm$0.13 mag (100 day)$^{-1}$ in the $B$ and 1.83$\pm$0.11 mag (100 day)$^{-1}$ in the $V$. 
On the other hand, the $R$ and $I$ light curves show a flattening with the
slope being 1.51$\pm$0.05 mag (100 day)$^{-1}$ in $R$ and 1.70$\pm$0.06 mag (100 day)$^{-1}$ in
$I$. The steepening of the $B$ and $V$ light curves could be an indication of
early dust formation.

 A comparison of late time decline
rate of SN 2011dh with that of SN 2008ax shows that except for the $I$-band,
in which SN 2008ax declines faster, the late phase decline rates of SN 2011dh
and SN 2008ax are very close \citep{taubenberger11}. Using the available data
of SN 1993J, the late phase decline rate, during 100 - 300 days past explosion,
is estimated as 1.39, 1.71, 1.49 and 1.87 mag (100 day)$^{-1}$ in $B$, $V$, $R$
and $I$ bands, respectively. This indicates SN 2011dh declines faster than
SN 1993J (see also Figure \ref{fig_lcomp2}).


\subsubsection{Colour curves}

\begin{figure}
\resizebox{\hsize}{!}{\includegraphics{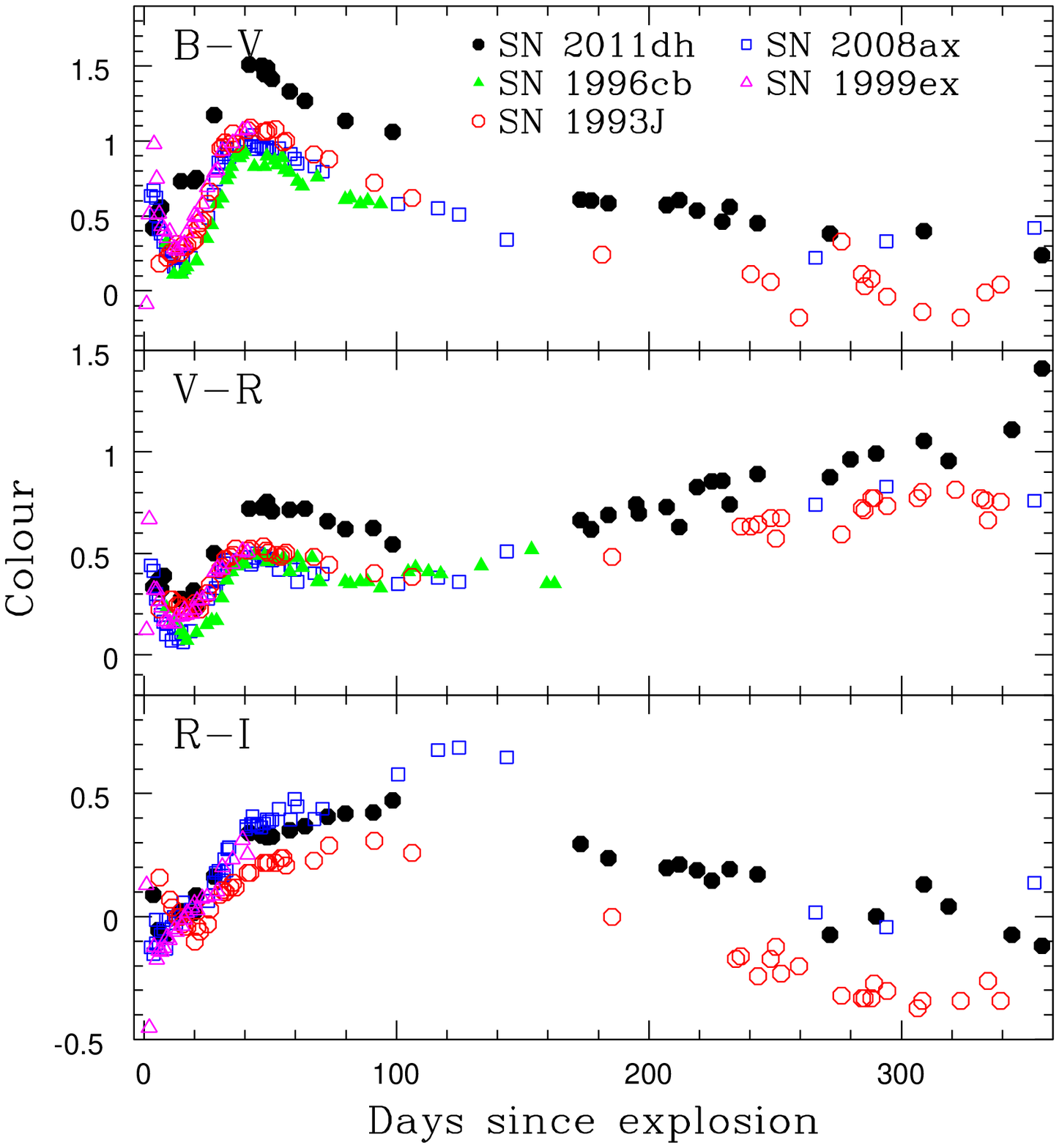}}
\caption[]{Comparison of colour curves of SN 2011dh with those of 
SN 2008ax, SN 1999ex, SN 1996cb and SN 1993J. \\
(A colour version of this figure is available in the online journal.)}
\label{fig_colourcomp}
\end{figure}

The colour curves of SN 2011dh along with supernovae 2008ax, 1996cb, 1993J and 
1999ex have been plotted in Figure \ref{fig_colourcomp}. The colour curves of 
SNe 2011dh, 2008ax, 1996cb, 1993J and 1999ex have been corrected for
$E(B-V)$ of 0.035, 0.4, 0.12, 0.18 and 0.30, respectively. It is evident from 
Figure \ref{fig_colourcomp} that the reddening corrected $(B-V)$ and $(V-R)$ 
colours of SN 2011dh are always redder as compared to the other stripped 
envelope core-collapse supernovae used in comparison. A  similar trend is seen 
in the $(R-I)$ colour also, though not very significant. An additional reddening
$E(B-V)$ of $\sim 0.35$ is required to bring the $(B-V)$ and $(V-R)$ colours of 
SN 2011dh close to those of other supernovae. High resolution spectroscopy of 
SN 2011dh does not reveal presence of any absorbing system within the host 
galaxy, implying insignificant reddening within the host galaxy. The other 
possibilty is that supernova SN 2011dh is intrinsically redder. A blackbody
fit to the spectrum obtained $\sim 3$ days after explosion gives a temperature 
of $\sim$ 7600 K \citep{arcavi11}, which is lower than that expected, at similar
epochs, from the explosion of a Red Super Giant. \cite{arcavi11} have argued 
that the sharp decline in the {\it g} band light curve during the first two 
days after explosion also needs a much lower temperature. 
It thus appears that the observed redder colour of SN 2011dh is due to its 
lower temperature and not because of excess reddening within the host galaxy.
      
\subsubsection{Absolute magnitude, bolometric light curve and mass of $^{56}Ni$ }
\begin{figure}
\resizebox{\hsize}{!}{\includegraphics{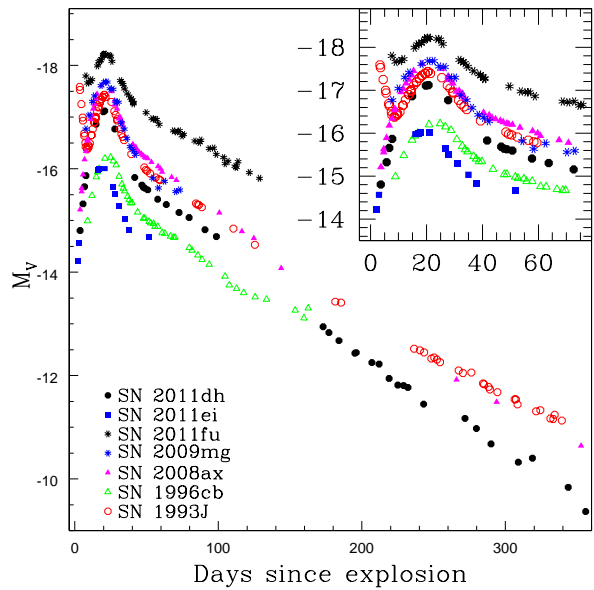}}
\caption[]{Comparison of the absolute $V$ light curve of SN 2011dh with those
of SN 2011fu, SN 2011ei, SN 2009mg, SN 2008ax, SN 1996cb and SN 1993J. Inset 
shows the early evolution of the light curve.\\
(A colour version of this figure is available in the online journal.)}
\label{fig_absmag}
\end{figure}

The $V$-band peak absolute magnitudes of SN 2011dh  estimated
adopting  reddening $E(B-V)$ = 0.035 mag  and  distance  8.4$\pm$0.7 Mpc \citep{vinko12} is 
$-17.123\pm0.18$ mag. The peak absolute magnitude of SN 2011dh is $\sim$ 1 mag fainter than the 
mean peak absolute magnitudes of the entire sample of stripped-envelope CCSNe and is $\sim$ 0.3 mag 
fainter than the type IIb sample \citep{richardson06}.  
It is fainter than the absolute peak $V$ magnitude of some other well studied type IIb supernovae 
 SN 1993J ($-17.57\pm0.24$, \citealt{maund04}), SN 2003bg (-17.50, \citealt{hamuy09}; \citealt{mazzali09}), 
SN 2008ax ($-17.617\pm0.43$, \citealt{taubenberger11}), SN 2009mg ($-17.68\pm0.48$, \citealt{oates12})
and SN 2011fu ($-18.50\pm0.24$, \citealt{kumar12}). On the other hand, the 
absolute peak  $V$-band  magnitude of SN 2011dh is brighter than SN 1996cb 
(-16.22, \citealt{qui99}), 
SN 2007Y ($-16.45\pm0.6$, \citealt{stritzinger09}) and SN 2011ei ($\sim -16$, 
\citealt{milisavljevic12}). It shows 
that type IIb supernovae represent an inhomogeneous class in terms of peak 
$V$-band magnitude, with a wide spread, of more than 2 mag. The absolute 
$V$-band light curve of SN 2011dh is plotted along with some well studied 
stripped envelope core collapse supernovae in Figure \ref{fig_absmag}. 
 The light curves indicate that SN 2011dh
declines at a rate of 1.83 mag (100 day)$^{-1}$ during days 172--353, while 
SN 2008ax declines at a rate of 1.66 mag (100 day)$^{-1}$ during days 144--353,
and SN 1993J declines at a rate of 1.46 mag (100 day)$^{-1}$ during days 
181--340. It appears that, in general, the light curves of type IIb SNe decline
with a rate faster than the decline rate expected from the decay of 
$^{56}$Co $\rightarrow$ $^{56}$Fe (0.98 mag (100 day)$^{-1}$), indicating the 
$\gamma$-rays produced in the decay may not be completely trapped by the 
supernova ejecta.
 
\begin{figure}
\resizebox{\hsize}{!}{\includegraphics{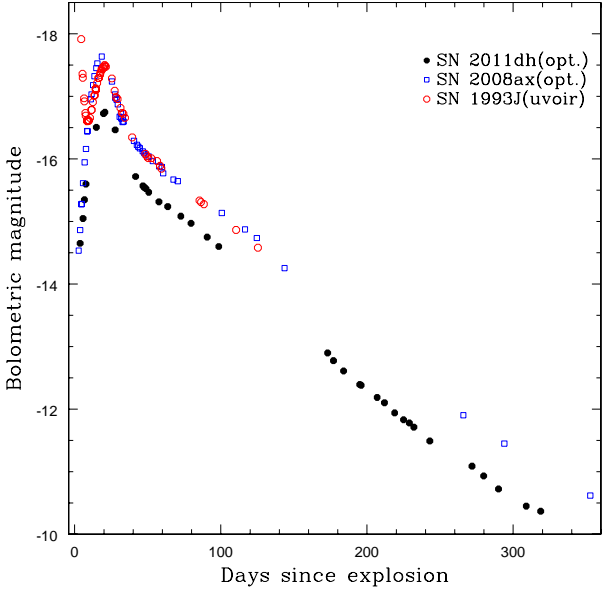}}
\caption[]{Quasi bolometric light curve  of SN 2011dh. Also plotted in the figure, for
comparison, are the bolometric light curves of SN 1993J and  SN 2008ax.\\
(A colour version of this figure is available in the online journal.)}
\label{fig_bol}
\end{figure}

Quasi bolometric light curve of SN 2011dh is obtained using the observed 
$UBVRI$ magnitudes corrected for reddening, and converted to their respective 
monochromatic flux using the zero points provided by \cite{bessell98}. The 
bolometric fluxes are derived by fitting a spline curve to the $U, B, V, R$ and 
$I$ fluxes and integrating over the wavelength range 3100~\AA\ to 10600~\AA\, 
determined by the response of the filters used for observations. We have 
observations in the $U$-band till November 19, and hence the contribution from 
$U$-band is included in the bolometric light curve till November 19 only. There 
are a few nights when we do not have observations in either the $U$, $B$ or $I$ 
bands. For estimating bolometric flux, the magnitudes of the missing bands, on 
these nights, were estimated by interpolating the observed magnitudes of the 
neighbouring nights. The quasi bolometric light curve of SN 2011dh alongwith 
the bolometric light curves of type IIb  SN 2008ax and SN 1993J  is plotted in 
Figure \ref{fig_bol}. The quasi bolometric light curves of SN 2008ax 
is  constructed using the published $U, B, V, R$ and $I$ magnitudes, 
(\citealt{pastorello08},  \citealt{taubenberger11}) in a manner similar to 
SN 2011dh. The bolometric light curve of SN 1993J was taken from  
\cite{lewis94}, which includes optical and NIR data. The total reddening for 
SN 2008ax is taken as $E(B-V)$=0.4 mag \citep{taubenberger11} and  distance of 
9.6 Mpc \citep{pastorello08} is used. The rise to maximum and subsequent 
evolution of the quasi bolometric light curve of SN 2011dh is similar to the 
other type IIb supernovae in comparison. However, SN 2011dh is fainter than 
both SN 2008ax and SN 1993J. 

The mass of $^{56}$Ni required to power the quasi bolometric light curve can be 
estimated using Arnett's rule \citep{arnett82}. Under the assumption that the 
radioactivity powers the light curve and at maximum light, most of the energy 
released by radioactivity is still being trapped and thermalized, the peak 
radiated luminosity is comparable to instantaneous rate of energy release 
of radioactive decay of $^{56}$Ni  synthesized during the explosion. The 
simplified form of Arnett's rule is expressed as 
\begin{equation}
M_{Ni} = L_{bol}/\alpha S(t_{R}),
\end{equation}
where, $\alpha$ is the ratio of bolometric to radioactivity luminosity (near 
unity) and $S$ is the radioactivity luminosity per unit nickel mass, evaluated 
at rise time t$_{R}$ \citep{nugent95}. In the case of SN 2011dh, the explosion 
date has been well constrained to better than 0.6 days between the first
detection on May 31.893 and last non-detection on May 31.275 \citep{arcavi11}. 
The date of explosion can be taken as a mean of these two dates. The photometry 
reported in section 3.1.1 and Table \ref{tab_parameter} shows that the supernova
reached maximum in $B$-band on JD 245\,5732.60, with $B$-band rise time of 
$\sim$ 19.6 days. The peak bolometric luminosity of SN 2011dh, estimated using 
the observed $UBVRI$ flux is 1.267$\times$ 10$^{42}$ erg sec$^{-1}$. Taking  
the  $B$-band  rise time of SN 2011dh as 19.6  days, mass of $^{56}$Ni  
synthesized in the explosion is estimated to be  0.063$\pm$0.011 M$\odot$. 

 The mass of $^{56}$Ni synthesized during the explosion, can  also be 
estimated by fitting the energy deposition rate via the
$^{56}$Ni$\rightarrow$$^{56}$Co$\rightarrow$$^{56}$Fe chain, to the early 
post-maximum observed bolometric light curve. The energy deposition rate for 
different values of $^{56}$Ni via the 
$^{56}$Ni$\rightarrow$$^{56}$Co$\rightarrow$$^{56}$Fe chain is estimated
using the analytical formula by \cite{nadyozhin94}. It is found that the energy 
deposition rate corresponding to mass of $^{56}$Ni as 0.070 M$\odot$ fits the early 
post-maximum decline of the observed quasi bolometric light curve of SN 2011dh.

 \cite{vinko04} provide a simple analytic model to fit the observed
bolometric light curve of supernovae. This model takes into account the energy 
deposition due to $\gamma$-rays produced in the decay chain 
$^{56}$Ni$\rightarrow$$^{56}$Co$\rightarrow$$^{56}$Fe and due to positrons.
We fit the early post-maximum decline phase ($<$ 30 days after maximum) of the 
observed bolometric light curve following the formulation by Vinko et al., to
estimate the mass of $^{56}$Ni. 
For the total ejected mass of $\sim$ 2 M$\odot$ \citep{bresten12} and the expansion velocity 
of the ejecta as inferred from Fe\,II lines,  
the optical depth for $\gamma$-rays and positrons during the early post-maximum decline phase is 
found to be high and the probability of $\gamma$-rays and positrons to escape 
from the ejecta  is very
low. Under the assumption that the diffusion time is short, mass of $^{56}$Ni is estimated by 
approximating the energy emitted per second (i.e. bolometric luminosity)  to the rate of the 
energy deposition at different times. A $^{56}$Ni mass of 0.067$\pm$0.013 
M$_\odot$ is found to fit the early post-maximum decline of the quasi 
bolometric light curve.

 The light curve of supernovae at late phases is powered by the 
$^{56}$Co $\rightarrow$ $^{56}$Fe decay. If all the $\gamma$-rays produced by the decay
are trapped in the supernova ejecta, the exponential tail
of the bolometric light curve during  the nebular phase can also be used to constrain the mass of  
$^{56}$Ni produced in the explosion \citep{hamuy03}. The slope of the bolometric 
light curve of SN 2011dh during days 60 to 100 is 1.79 mag (100 day)$^{-1}$, 
which is faster compared to the expected decay rate of 
$^{56}$Co$\rightarrow$ $^{56}$Fe (0.98 mag (100 day)$^{-1}$). However, the
observed slope can be used to estimate a lower limit of the mass of $^{56}$Ni.
Using the bolometric flux during days 60 to 98 we calculate the lower limit of 
$^{56}$Ni mass as 0.037$\pm$0.004 M$\odot$.
  
It is to be noted here that while calculating the bolometric light curve of 
SN 2011dh, contribution due to missing bands in ultra-violet and infra-red has 
not been taken into account. In the case of SN 1993J \cite{wada97} have shown 
that before the second maximum, the flux emitted in $J$-band was $\sim$ 8\% of 
the total flux from $U$ to $J$-band, and was $\sim$ 15\% after the second 
maximum. The black body fit to the photometric data shows that the photospheric 
temperature of SN 1993J close to maximum light was $\sim 8200$ K 
\citep{lewis94}.  \cite{richmond94}  have estimated the fraction of total 
blackbody flux emitted in the $UBVRI$ bands and shown it to be $\sim 70$\% 
at a blackbody temperature of $\sim$ 8000 K. For SN 2008ax, the UV contribution 
to the pseudo-bolometric light curve is always less than $\sim$ 15\%, 
and at the time of maximum it is less than $\sim$ 10\%  \citep{taubenberger11}. 
 In a recent paper, \cite{marion13} show that in the case of SN 2011dh
the NIR contribution at peak is $\sim$ 35\%, and increases to 
$\sim$ 52\% by day 34. After accounting for the missing
NIR band flux, the peak bolometric flux of SN 2011dh is 1.711$\times$ 10$^{42}$  erg sec$^{-1}$
and the mass of $^{56}$Ni is 0.084 M$\odot$. The inclusion of  missing  NIR flux to quasi 
bolometric flux leads to  $^{56}$Ni  mass of 0.095 M$\odot$ and 0.091 M$\odot$, using  
\cite{nadyozhin94} and \cite{vinko04} formulation, respectively.  

\subsection{Spectroscopic results}
\subsubsection{Pre-maximum spectral evolution}

\begin{figure}
\resizebox{\hsize}{!}{\includegraphics{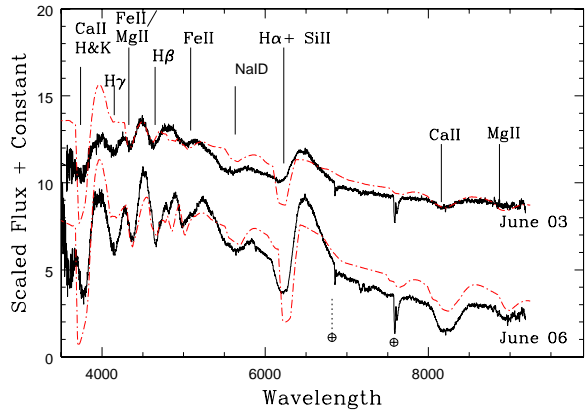}}
\caption[]{Pre-maximum spectral evolution of SN 2011dh during $\sim$ 3 to $\sim$ 6 days after explosion.
The observed spectra (black, continuous) have been plotted with the synthetic 
spectra (red, dash-dotted) created using SYN++. \\
(A colour version of this figure is available in the online journal.)}
\label{fig_early_spec1}
\end{figure}

The pre-maximum spectra of SN 2011dh obtained $\sim$ 3 and 6 days after 
explosion are shown in Figure \ref{fig_early_spec1}. The spectrum of day 3 shows
a blue continuum that drops below $\sim 5000$~\AA . The prominent noticable 
features in the first spectrum are the broad P-Cygni absorption of H$\alpha$ 
and absorptions due to H$\beta$ and H$\gamma$, Ca\,II H\&K and  NIR triplet. To 
identify the other features seen in the spectrum, the  observed spectrum is 
compared  with the synthetic spectrum generated using SYN++. 

SYN++ is a rewrite of the parameterized spectrum synthesis code SYNOW, with a 
new structured input control file format  and with  more complete atomic data 
files \citep{thomas11}. The basic assumption of SYNOW includes spherically 
symmetric and homologously expanding ejecta $(v \propto r)$, local 
thermodynamic equilibrium (LTE) for level populations and resonant scattering 
line formation above a sharp  photosphere emitting a blackbody continuum. The 
line formation is treated using the Sobolev approximation (\citealt{sobolev57}, 
\citealt{jeffery90}). The optical depth $\tau$ of the strongest line is a free 
fitting parameter and optical depths of other lines of the same ion are 
determined assuming Boltzmann equilibrium at excitation temperature $T_{exc}$.  
A detailed description of SYNOW has been presented in \cite{fisher00} and 
\cite{branch02}.

The synthetic spectra for days 3 and 6 are plotted with the observed spectra
in Figure \ref{fig_early_spec1}. The main species required to produce the 
observed features have been marked. The synthetic spectrum of day 3 has a 
blackbody continuum temperature $T_{bb}=7500$ K, similar to \cite{arcavi11}, 
and photospheric velocity $v_{\rm{ph}}=12000$ km sec$^{-1}$. The hydrogen line is 
detached from the photosphere with a minimum velocity $v_{\rm{min}}=15000$ km 
sec$^{-1}$. All 
other lines are undetached. The H$\beta$ and H$\gamma$ absorptions are fit 
reasonably well with $v_{\rm{min}}=15000$ km sec$^{-1}$ and 
$v_{\rm{max}}= 23000$ km sec$^{-1}$. However, it is difficult to fit the broad 
absorption of H$\alpha$ probably due to the LTE and resonance scattering 
assumptions of SYN++, or due to blend of H$\alpha$ with other features. 
\cite{baron95} in their study of SN 1993J attribute the broadness of the 
H$\alpha$ absorption to a blend of H$\alpha$ with Fe\,II lines, while 
\cite{elmhamdi06} suggest it to be a combination of H$\alpha$ and Si\,II in the
case of SN 2000H. The broad H$\alpha$ profile, on the other hand, is 
interpreted as due to two components of hydrogen with different velocities in 
the case of SN 2011ei by \cite{milisavljevic12}. We have explored the 
possibility of two components of hydrogen to fit the broad H$\alpha$ absorption,
and find that it does not fit well.  A combination
of H$\alpha$ and Si\,II also does not fit this feature well (see Figure 
\ref{fig_early_spec1}), indicating the requirement of a more detailed
modelling. The prominent absorption  feature seen at 
$\sim$ 5500~\AA\ is usually attributed to He\,I 5876~\AA\ and Na\,ID 5890, 5896
\AA\ lines (\citealt{pastorello08}, \citealt{milisavljevic12}). Since we do not 
find signature of other lines due to He\,I, the absorption is most likely due 
to Na\,ID alone.        
 
The spectrum of day 6 (6 June) is very similar to the first spectrum. The 
absorption lines have become sharper and stronger. A synthetic spectrum, with 
a photospheric temperature T$_{bb}$ = 6500 K and photospheric velocity 
$v_{ph}$ = 10000 km sec$^{-1}$  matches the observed spectrum well. In this 
case also, hydrogen line is detached from the photosphere with a minimum 
velocity $v_{min}$ =  12,000 km sec$^{-1}$. The same species used to produce 
the synthetic spectrum corresponding to the day 3 spectrum have been used. 

\subsubsection{Early post-maximum spectral evolution}

\begin{figure}
\resizebox{\hsize}{!}{\includegraphics{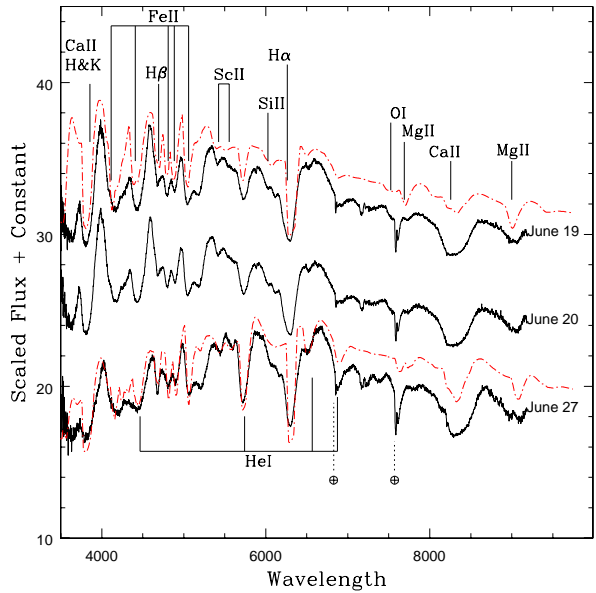}}
\caption[]{Spectral evolution of SN 2011dh during $\sim$ 20 to $\sim$ 27 days after explosion.
The observed spectra (black) have been plotted with the synthetic spectra (red, dash-dotted) 
created using SYN++.\\
(A colour version of this figure is available in the online journal.)}
\label{fig_early_spec2}
\end{figure}

We had a big gap in spectral monitoring and our next spectrum could be obtained 
only on 2011 June 19, corresponding to the maximum in $B$-band, $\sim 19.3$ 
days after the explosion. This spectrum together with two other spectra of 
nearby epochs have been shown in Figure \ref{fig_early_spec2}. In the spectrum 
of 19 June, a notch is seen in the emission component of H$\alpha$, giving it
a double peaked appearance. This is due to the emergence of the He\,I 6678~\AA\
feature. The P-Cygni features from other He\,I lines 5876~\AA\ and 7065~\AA\ 
are clearly identified in this spectrum. The expansion velocity of H$\alpha$ is 
$\sim$ 12800 km sec$^{-1}$ and that of He\,I 5876~\AA\  is $\sim 7300$ km 
sec$^{-1}$. The spectrum indicates the supernova has already entered into a 
phase wherein the lines due to He\,I become prominent. The onset of this phase 
occurred sometime between 6 June and 19 June. \cite{marion11} have reported 
non detection of helium lines in the NIR spectrum of SN 2011dh obtained on 
8 June, weak evidence of helium in the spectrum obtained on 12 June and an
unambiguous detection of helium lines at 10800~\AA\ and 20581~\AA\ in the 
spectrum obtained on June 16. A weak evidence of He\,I 5876~\AA\ and 6678~\AA\
lines was also indicated by them in the optical spectrum of 12 June, while
these lines were clearly seen in their spectrum obtained on 14 June. Thus,
the He\,I lines  developed in the spectrum of SN 2011dh $\sim$ 13 days after 
the explosion. This is similar to SN 2008ax, in which the He\,I lines are found 
to emerge in the NIR spectra  at $\sim$ 11 days after explosion \citep{taubenberger11}.

The next two spectra obtained on 20 June and 27 June are very similar to that 
of 19 June. The He\,I lines become stronger. In the spectrum of 27 June, 
[O\,I] 5577~\AA\ line starts appearing. 

The spectra of 19 June and 27 June were modelled with the SYN++ code. The 
synthetic spectra have been plotted with the observed spectra in Figure 
\ref{fig_early_spec2}. The spectrum of 19 June is computed with 
$T_{\rm{bb}}= 6500$~K and $v_{\rm{ph}} = 7000$ km sec$^{-1}$, while that of 
27 June is computed with  
$T_{\rm{bb}}=4800$~K and  $v_{\rm{ph}} = 6300$ km sec$^{-1}$. The main features 
have been identified and are marked in the figure. Hydrogen and He both are 
detached from the photosphere with $v_{\rm{min}}$ of 10,000~km sec$^{-1}$ and 
7500 km sec$^{-1}$, respectively. The broad absorption of H$\alpha$ is still 
not well reproduced, however, Si\,II helps to improve the fit. H$\beta$ line is 
fit well, H$\gamma$ gets blended with Fe\,II lines.  Other narrow Fe\,II lines 
are well reproduced.  Introduction of Sc\,II (lines at 5527~\AA \ and 5661~\AA) 
is required to produce the two weak troughs blueward of He\,I 5876~\AA\ 
absorption. In the red region of the spectrum, lines due to O\,I,  Mg\,II and  
Ca\,II are well reproduced.      

\begin{figure}
\resizebox{\hsize}{!}{\includegraphics{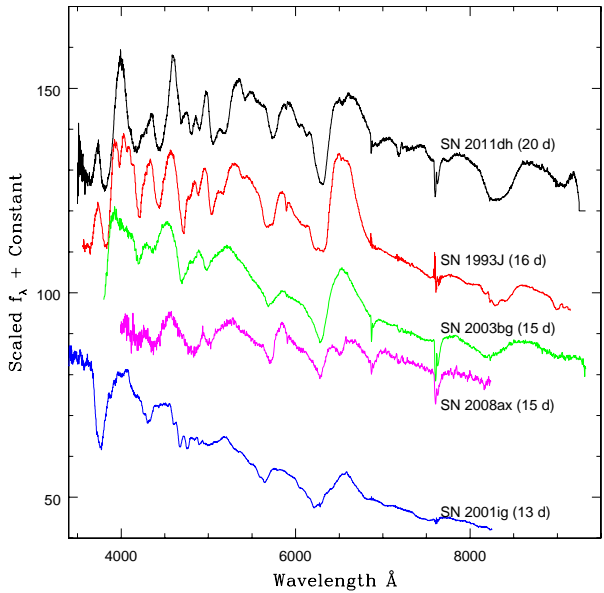}}
\caption[]{Comparison of  spectra of  SN 2011dh with those of SN 1993J, SN 2003bg, SN 2008ax and SN 2001ig
during the early (close to maximum in $B$-band) phase.\\
(A colour version of this figure is available in the online journal.)}
\label{fig_early_spec_comp}
\end{figure}

Figure \ref{fig_early_spec_comp} shows the spectrum of SN 2011dh close to $B$
maximum, together with the spectra of other type IIb supernovae (obtained from 
SUSPECT archive) at a similar epoch. The lines due to helium are seen in the 
spectrum of all the supernovae, although with varying strength. The He\,I 
5876~\AA\ line is well developed in all the supernovae. He\,I 6678~\AA\ has 
just started appearing in SN 2011dh and SN 1993J, it is stronger in SN 2008ax,
while it is not seen in the spectrum of SN 2003bg and SN 2001ig. There is a 
strong  similarity between  spectra of SN 2011dh and SN 1993J, the strengths 
of H$\alpha$ and He\,I 5876~\AA\ are similar in both the objects. SN 2011dh 
has the strongest Ca\,II NIR feature.
At $\sim$ 30 days after explosion, the strengths of H$\alpha$ and the He\,I 5876~\AA\ line 
were comparable in the spectrum of SN 2008ax \citep{pastorello08}, 
while, H$\alpha$ is found to be stronger than He\,I 5876 in the spectrum of 
SN 2011dh at $\sim$ 27 days after explosion.  

\subsubsection{Transitional phase}

\begin{figure}
\resizebox{\hsize}{!}{\includegraphics{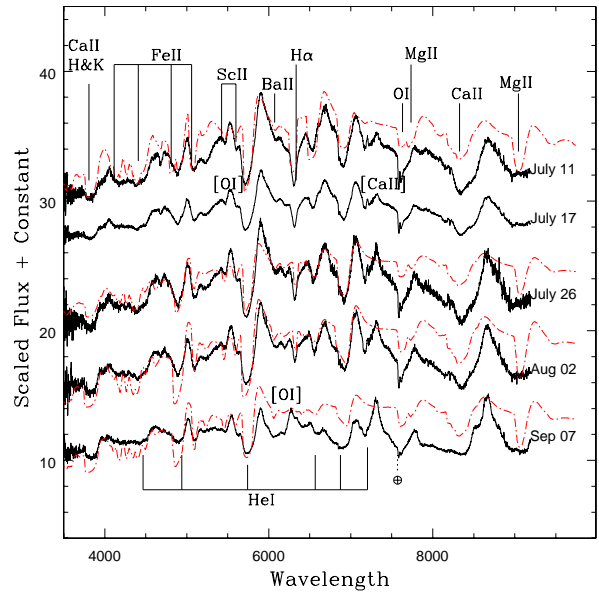}}
\caption[]{Spectral evolution of SN 2011dh during 41 to 99 days after explosion.
The observed spectra (black) have been plotted with the synthetic spectra (red, dash-dotted) 
created using SYN++.\\
(A colour version of this figure is available in the online journal.)}
\label{fig_postmax_spec}
\end{figure}

The spectra of SN 2011dh during $\sim 41$ to $\sim 99$ days after explosion 
are shown in Figure \ref{fig_postmax_spec}. A considerable evolution is seen
in the spectrum during this period. The continuum becomes redder, the Balmer 
lines become sharper and have lower expansion velocities. The lines due to 
He\,I become stronger. The metamorphosis of SN 2011dh spectrum from type II 
like to type Ib like takes place during this period. The overall shape of the 
spectrum obtained on 11 July (day 41) is similar to the one obtained on 27 June.
The lines due to He\,I become stronger, P-Cygni features due to  He\,I lines 
4471, 5015, 5876, 6678, 7065 and 7281 \AA\ are well developed. H$\alpha$ line 
is weaker and narrower. The absorption components due to Ca\,II H\&K and Ca\,II 
NIR triplet weaken, and  the emission component of Ca\,II NIR becomes stronger. 
The [Ca\,II] 7291, 7324~\AA\ feature starts appearing, blended with 
He\,I 7281~\AA. The forbidden line [O\,I] at 5577~\AA\ is also prominent. 

 The spectral evolution between 11 July to 2 August is very slow. The 
H$\alpha$ absorption continues to weaken, the strength of the [Ca\,II] 
feature and Ca\,II NIR triplet increases. Except for the narrow H$\alpha$ 
absorption, the spectrum of SN 2011dh displays all the features of type 
Ib supernova. The emergence of forbidden lines due to calcium and oxygen 
shows that the spectrum is entering into nebular phase.  

The spectrum of 7 September shows a complete transformation from type IIb
to type Ib. The [O\,I] feature at 6300, 6363 is seen, while the H$\alpha$ 
absorption has completely disappeared. The  He\,I lines at 5876, 6678 and 7065 
start weakening, the absorption component of  Ca\,II NIR triplet becomes weaker,
and the spectrum appears to be dominated by emission lines. 
   
Synthetic spectra modelled with the SYN++ code is also plotted in Figure 
\ref{fig_postmax_spec}, alongwith the observed spectra. There is not much 
evolution seen in the photospheric velocity and the photospheric temperature 
during this phase. The synthetic spectrum of 11 July has $T_{\rm{bb}}=4300$~K 
and $v_{\rm{phot}}=4500$~km sec$^{-1}$. Hydrogen is still detached from the 
photosphere with  $v_{\rm{min}} = 10,500$~km sec$^{-1}$, whereas helium is now 
undetached. The absorption blueward of H$\alpha$ can be fit with Ba\,II 6142
and 6496~\AA\ lines. The synthetic spectrum which fits the observed spectrum of
7 September has a very similar blackbody temperature and photospheric velocity
as that of 11 July, but no hydrogen.

\begin{figure}
\resizebox{\hsize}{!}{\includegraphics{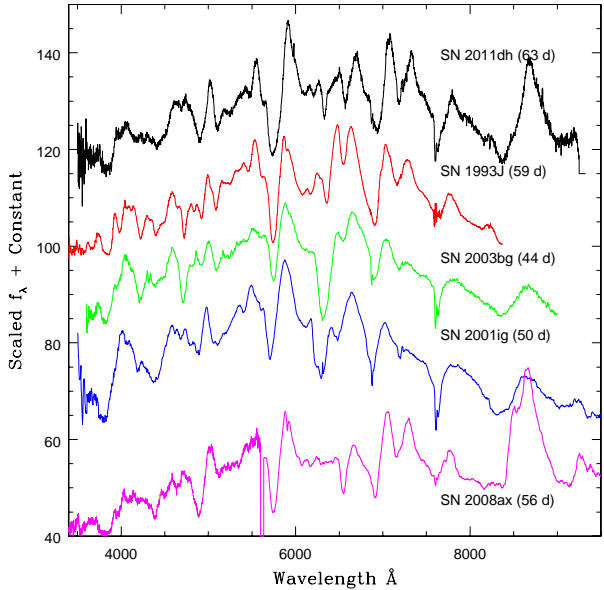}}
\caption[]{Comparison of the spectrum of SN 2011dh with those of SN 1993J, 
SN 2003bg, SN 2001ig and SN 2008ax at $\sim$ 2 months after explosion.\\
(A colour version of this figure is available in the online journal.)}
\label{fig_postmax_spec_comp}
\end{figure}

The spectrum of SN 2011dh $\sim 60$ days after explosion is compared with other
type IIb supernovae in Figure \ref{fig_postmax_spec_comp}. The overall 
appearance of the day 63 spectrum of SN 2011dh is very similar to the day 59 
spectrum of SN 1993J. In the spectra of all the objects shown in the Figure, 
the He\,I 5876~\AA\ absorption is strong, but the absorption due to H$\alpha$ 
has a varying strength. In SN 2011dh and SN 1993J the H$\alpha$ absorption is
narrow and sharp, SN 2003bg and SN 2001ig still show broad, strong absorption,
and in SN 2008ax it is almost negligible. In SN 2009mg, \cite{oates12} have 
shown that the H$\alpha$ absorption was considerably strong $\sim$ 60 days 
after explosion, and completely disappeared by $\sim$ 110 days spectrum. In 
SN 1996cb a narrow H$\alpha$ absorption was seen till day 107, and nearly 
vanish by day 114 \citep{qui99}.   

The transition from a hydrogen dominated early phase spectra to He dominated 
late phase spectra is the characteristic feature of type IIb supernovae. This 
can be interpreted as the presence of a thin envelope of hydrogen at the time 
of explosion of the progenitor star. However, it appears that the  exact amount 
of hydrogen the progenitors retain at the time of explosion differs 
significantly from object to object  and  mass of H envelope may play 
a key role in the photometric and spectroscopic evolution of these objects. 
The observed features of the light curve of SN 1993J are well reproduced by 
the explosion of a red supergiant whose H/He envelope mass has been decreased 
below $\sim$ 0.9 M$_\odot$ \citep{shigeyama94}. \cite{woosley94} arrived at the 
mass of hydrogen envelope as 0.2$\pm$0.05 M$_\odot$. Using the late time 
spectra, \cite{houck96} estimated $\sim$ 0.3 M$_\odot$ as mass of hydrogen 
envelope in SN 1993J. \cite{mazzali09} have estimated mass of hydrogen layer 
as 0.05 M$_\odot$ for the broad-line type IIb supernova SN 2003bg. In the case
of SN 2008ax, \cite{chornock11} have estimated the mass of hydrogen envelope as 
$\sim$ few$\times$ 0.01 M$\odot$. The overall similarity of the spectral 
evolution of SN 2011dh with SN 1993J indicates that the mass of the hydrogen 
envelope in SN 2011dh could be similar to that of SN 1993J. Indeed, 
\cite{bresten12} have shown that a progenitor star with He core of 3 to 4 
M$_\odot$ with a  thin hydrogen envelope of $\sim$ 0.1  M$_\odot$ can lead to a 
light curve  similar to SN 2011dh.  

\subsubsection{Nebular phase}

\begin{figure}
\resizebox{\hsize}{!}{\includegraphics{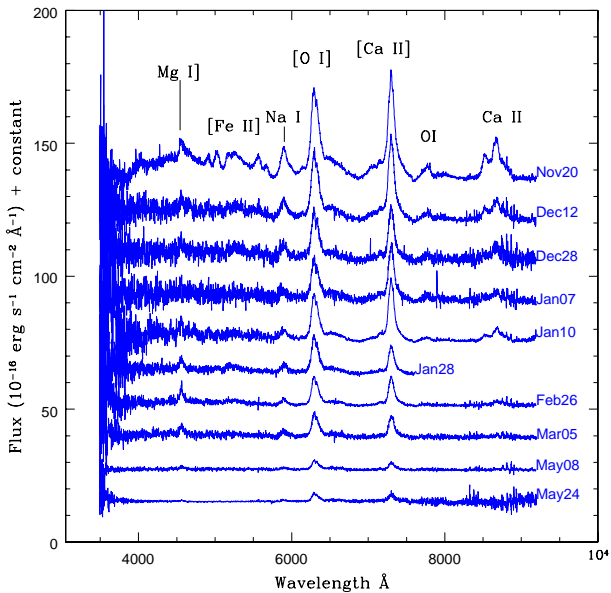}}
\caption[]{Spectral evolution of SN 2011dh during $\sim$ 173 to $\sim$ 360 days after explosion.\\
(A colour version of this figure is available in the online journal.)}
\label{fig_neb_spec}
\end{figure}

The next set of spectra were obtained when the supernova reappeared in the 
night sky, on ten occasions, during 2011 November 20 to 2012 May 24. These 
spectra are shown in Figure \ref{fig_neb_spec}.  The He\,I lines, which were
prominent in the spectra obtained until 2011 September 7, are totally absent. 
Instead, the spectra are dominated by emission lines of Mg\,I] 4571~\AA,
[O\,I] 6300, 6363~\AA, [Ca\,II] 7291, 7324~\AA, O\,I 7774~\AA, blend of [Fe\,II]
 lines at $\sim$ 5000~\AA\  and the Ca\,II NIR triplet. The [O\,I] 5577~\AA\ 
line had weakend, and almost disappeared in the spectrum of 2012 January 28.
A bump, redward of [O\,I] 6300, 6363~\AA, is seen in all the late phase spectra 
presented here.  

\begin{figure}
\resizebox{\hsize}{!}{\includegraphics{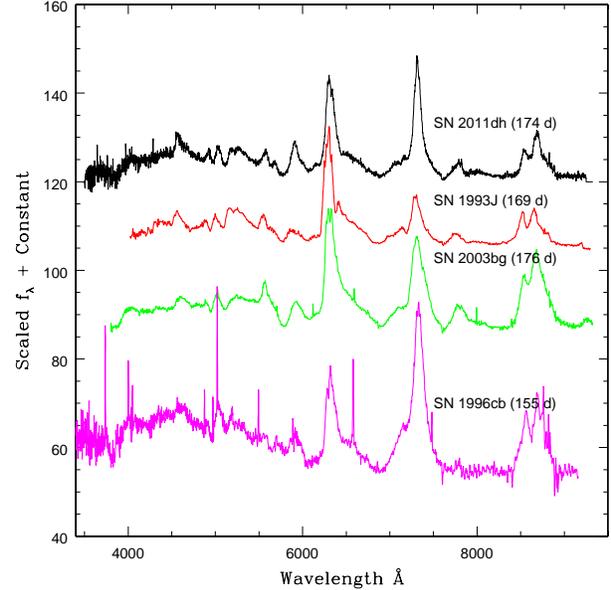}}
\caption[]{A comparison of the spectrum of SN 2011dh with those of SN 1993J, 
SN 2003bg and SN 1996cb at epoch $\sim$ 6 months after explosion.\\
(A colour version of this figure is available in the online journal.)}
\label{fig_spec_180day_comp}
\end{figure}

In Figure \ref{fig_spec_180day_comp}, the spectrum of SN 2011dh $\sim$ 6 months 
after explosion is compared with the spectra of SN 1993J, SN 2003bg and 
SN 1996cb at a similar epoch. Except for the difference in the strength of 
[Ca\,II] 7291, 7324~\AA\ lines, all the spectra are similar. The flux ratio 
of the [Ca\,II] 7291+7324 and [O\,I] 6300+6363 lines is highest ($\sim$ 2) 
in SN 1996cb and the least ($\sim$ 0.2) in SN 1993J, while it is intermediate
at a value of $\sim$ 0.8 and $\sim$ 0.5, respectively, in SN 2011dh and 
SN 2003bg. The bump redward of [O\,I]6300, 6363~\AA\ line is seen in all the 
objects. During the early nebular phase, the observed bump may be the result 
of contribution from Fe\,II, [Fe\,II] and possibly [Co\,II] \citep{patat95}, and 
at least some part of the flux could arise from H$\alpha$ scattering 
\citep{houck96} also.    

\begin{figure}
\resizebox{\hsize}{!}{\includegraphics{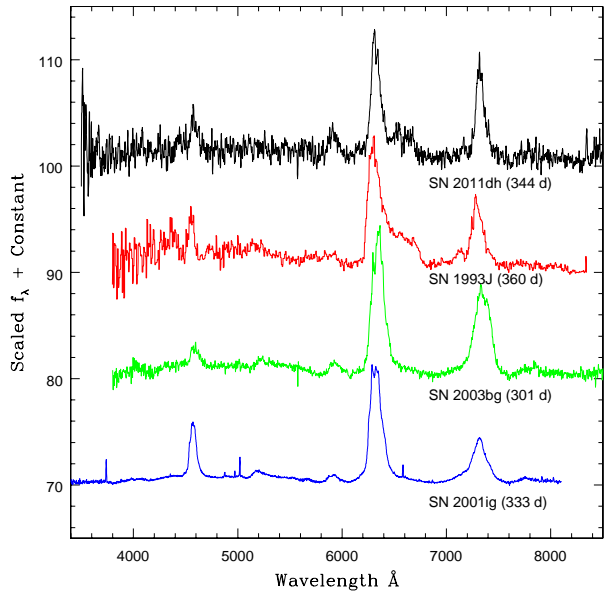}}
\caption[]{A comparison of the spectrum of SN 2011dh with those of SN 1993J, 
SN 2003bg and SN 2001ig at epoch $\sim$ 1 year after explosion.\\
(A colour version of this figure is available in the online journal.)}
\label{fig_neb_comp_spec}
\end{figure}

The spectrum of SN 2011dh obtained around $\sim$ 340 days after explosion is 
plotted alongwith the spectra of other well studied type IIb supernovae in 
Figure \ref{fig_neb_comp_spec}. The common nebular features of all the 
supernovae are similar, however, there are some differences with respect to the 
relative strength of emission lines and their shapes. The Mg\,I]/[O\,I] line 
ratio at $\sim$ 280 days after explosion is $\sim 0.2$ in SN 2011dh, 
$\sim 0.15$ in SN 1993J, and $\sim 0.36$ in SN 2001ig, which is known to have 
the strongest Mg\,I] feature \citep{silverman09}. The strength of H$\alpha$
 (redward of  [O\,I]6300, 6363~\AA\ line) has increased, with the feature having 
a box shaped profile, similar to that seen in SN 1993J, SN 2007Y and SN 2008ax. 

The exact mechanism producing the late time H$\alpha$ emission is not clearly 
known. However, the possible processes which can explain the late time H$\alpha$
emission are - ionization of hydrogen by radioactivity and/or by the X-rays 
emitted due to the ejecta-wind interaction. For different objects, either of 
the two mechanisms may be operational. In SN 1993J, this feature was 
interpreted as due to H$\alpha$ emission from a shell of hydrogen, possibly 
excited by interaction with a dense circumstellar medium (\citealt{patat95}, 
\citealt{houck96}). The observed late phase H$\alpha$ emission in SN 2007Y was 
attributed to shock interaction by \cite{stritzinger09}, but, \cite{chevalier10}
proposed that the circumstellar density of SN 2007Y was too weak to produce the 
observed H$\alpha$ luminosity through shock interaction, and hence 
radioactivity alone was the source for ionization of hydrogen. \cite{taubenberger11} have 
shown that the shock-wave interaction mechanism for late H$\alpha$ emission 
from SN 2008ax faces serious problem in explaining the observed shape and 
velocity of the nebular H$\alpha$ emission. An alternative mechanism in which 
a right combination of mixing and clumping of hydrogen and helium, ionized by  
radioactive energy deposition has been shown to be able to reproduce the 
observed H$\alpha$ emission in the late phase \citep{maurer10}.

Radio emission has been detected from SN 2011dh at a very early stage.  It was
detected at 86 GHz by the CARMA Radio Telescope $\sim$ 3 days after 
discovery \citep{horesh11} and at 22 GHz $\sim$ 14 days after discovery \citep{marti11}.  
This early phase radio emission is consistent with a non-thermal synchrotron 
self absorbed spectrum of optical photons \citep{soderberg12}. 
 SN 2011dh was monitored in the radio during late phases also with the 
VLBI.  Based on these data  a 
time averaged expansion velocity of the forward shock was estimated as 
$21000\pm 7000$~km sec$^{-1}$ \citep{bietenholz12}.
The observed X-ray emission from SN 2011dh gives a strong indication of the 
presence of circumstellar material, as the X-ray emission can be interpreted as 
due to the interaction of the blast wave with its surrounding circumstellar 
medium \citep{campana12}. Based on the evidence for the presence of 
circumstellar material, and the high expansion velocity of forward shock, it 
appears that shock-wave interaction may be the most plausible mechanism for the 
late H$\alpha$ emission in SN 2011dh. A detailed modelling is however required
to ascertain this possibility. 

\begin{figure}
\resizebox{\hsize}{!}{\includegraphics{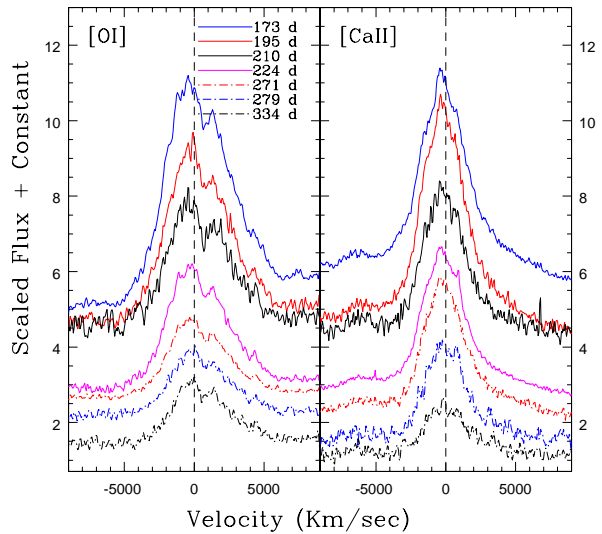}}
\caption[]{Line profile of [O\,I] $\lambda$$\lambda$ 6300, 6363 and [Ca\,II] $\lambda$$\lambda$7291, 7324 
lines during the nebular phase.\\
(A colour version of this figure is available in the online journal.)}
\label{fig_oxy_line_pro}
\end{figure}
 
The profile of the [O\,I] 6300, 6363~\AA\ and [Ca\,II] 7291, 7324~\AA\ lines 
during the nebular phase has been plotted in Figure \ref{fig_oxy_line_pro}. The [Ca\,II] line 
has a single peak, whereas the  [O\,I] line shows a double peak 
profile, with a $\sim$ 3000 km sec$^{-1}$ separation between the two peaks. Such
a double peaked [O\,I] profile, with a separation of $\sim$ 3000 km sec$^{-1}$ 
is seen in the late time spectra of many stripped-envelope
core-collapse supernovae (\citealt{taubenberger09}, \citealt{milisavijevic10}, \citealt{maeda08}), 
and is interpreted as due to asphericity in the explosion 
with preferred viewing angle, as spherical explosion can not produce the double 
peaked [O\,I] lines. However, \cite{maurer10} suggest that H$\alpha$ absorption 
is probably responsible for the double peak profile of the [O\,I] doublet in 
several type IIb supernovae.  They have shown that H$\alpha$ absorption causes 
a split of the [O\,I] doublet if it is located at around 12000 km sec$^{-1}$. The 
H$\alpha$ line velocity evolution in SN 2011dh (Figure \ref{fig_vel_evol}) 
shows that it starts with $\sim$ 17500 km  sec$^{-1}$ and flattens at 
$\sim$ 12000 km sec$^{-1}$  after day 25, which looks consistent with
the model suggested by \cite{maurer10}. 
 The spectra beyond day 173 show the emergence of H${\alpha}$ emission,
which could  be responsible for  redward asymmetry of [O\,I] profile (Figures 13, 14 and 15).
However, spectropolarimetric  observation and detailed modelling are required 
to differentiate between asphericity and H${\alpha}$  as the cause of the
double peak asymmetric profile.

\subsubsection{Expansion velocity of the ejecta}

\begin{figure}
\resizebox{\hsize}{!}{\includegraphics{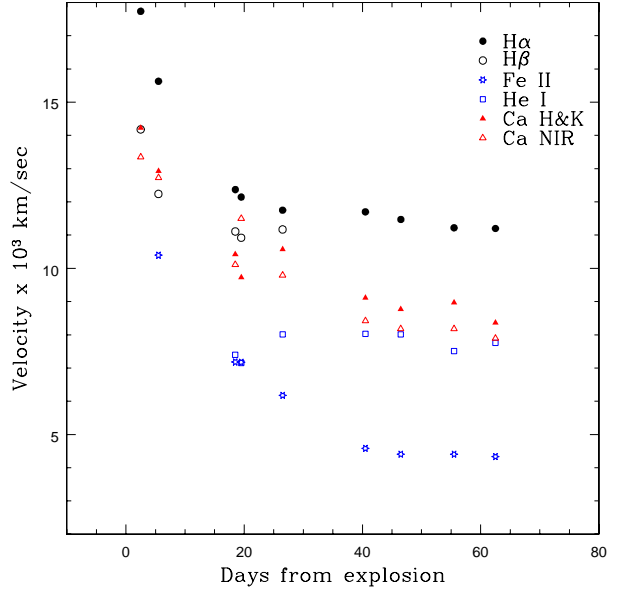}}
\caption[]{Evolution of the expansion velocity of supernova ejecta measured using
different species.\\
(A colour version of this figure is available in the online journal.)}
\label{fig_vel_evol}
\end{figure}

Expansion velocity of the ejecta has been measured from the minima of P-Cygni 
profiles of relatively isolated lines, {\it e.g.} H$\alpha$, H$\beta$, He\,I 
5876~\AA\ and Fe\,II 5169~\AA. For measuring the velocities, the absorption 
component of the P-Cygni profile was fitted with a Gaussian. The estimated line
velocities are plotted in Figure \ref{fig_vel_evol}. The expansion velocity of 
H$\alpha$ declines from $\sim$ 17500 km sec$^{-1}$ at $\sim$ 3 days 
after explosion to $\sim$ 12000 km sec$^{-1}$ $\sim$ 19 days after explosion, 
and then remains almost constant. The Ca\,II H\&K and Ca\,II NIR lines also
show a similar trend,  they  decline from an initial velocity of $\sim$ 14000 km 
sec$^{-1}$ to $\sim 10000$~km sec$^{-1}$, $\sim 19$ days after explosion, and
remains almost constant thereafter. The  expansion velocity of the ejecta 
measured using weak, unblended lines such as Fe\,II 5169~\AA, is often 
considered a good tracer of the photospheric velocity. The Fe\,II 5169 line 
velocity in SN 2011dh has an initial value of $\sim$ 10000 km sec$^{-1}$ that
declines until $\sim$ 40 days after explosion, and then flattens at 
$\sim$ 4500 km sec$^{-1}$. We do not find much evolution in the velocity 
of the He\,I 5876~\AA \ line, which shows a marginal increase from an initial
value of $\sim$ 7000 km sec$^{-1}$ to $\sim$ 8000 km sec$^{-1}$ and remains 
almost constant at this value. 

The measured expansion velocity  $\sim$ 20 days after explosion, shows three 
different representative velocities - the  H$\alpha$ line shows a higher 
velocity at $\sim$ 12000 km sec$^{-1}$, the Ca\,II H\&K and Ca\,II NIR triplet 
lines are at $\sim$ 10000 km sec$^{-1}$  and the Fe\,II line is at $\sim$ 7000 
km sec$^{-1}$. The velocity stratification becomes more prominent beyond day
40, with the H$\alpha$ velocity being around 11000 km sec$^{-1}$, the He\,I 
5876~\AA\ and the Ca\,II H\&K and Ca\,II NIR triplet at $\sim 8500$~km sec$^{-1}$
and the Fe\,II 5169~\AA\ line at a lower velocity of $\sim$ 4500 km sec$^{-1}$. 
Assuming the supernova ejecta expands homologously, the three velocity 
groups indicate the ejecta to have three layers. The outermost layer is made 
of a thin hydrogen layer moving with high velocity, the intermediate, denser 
layer consists of Ca and probably He, and the innermost high density core 
consisting of Fe and other heavy species  is moving with lower velocity. A
similar stratification of H and He layers is confirmed by spectral modelling 
of SN 2003bg \citep{mazzali09}, where the weaker Balmer lines level off at a 
velocity of $\sim$ 10000 km sec$^{-1}$ and He\,I lines level off at a lower 
velocity $\sim$ 7000 km sec$^{-1}$.

\begin{figure}
\resizebox{\hsize}{!}{\includegraphics{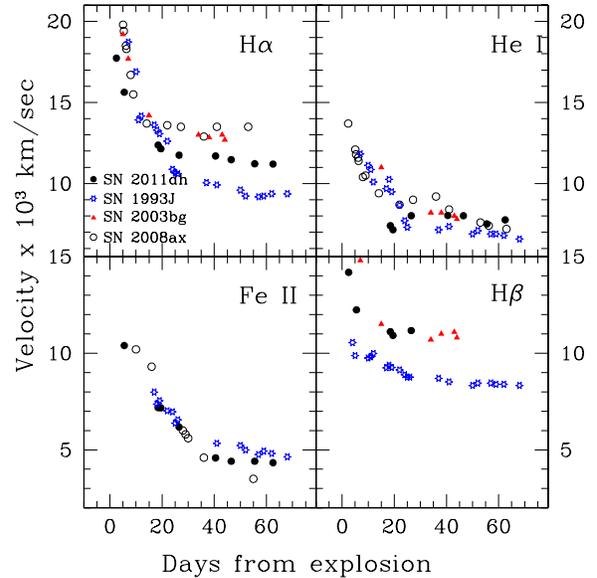}}
\caption[]{Comparison of the evolution of expansion velocity of SN 2011dh
with those of SN 1993J, SN 2003bg and SN 2008ax.\\
(A colour version of this figure is available in the online journal.)}
\label{fig_vel_evol_comp}
\end{figure}

In Figure \ref{fig_vel_evol_comp} the measured line velocities of SN 2011dh 
are plotted with those of SN 1993J, SN 2003bg  and SN 2008ax. The photospheric
velocity as measured using the Fe\,II line and the helium layer velocity are 
very similar in all the supernovae. However, differences are
seen in the Balmer line velocities. SN 2008ax and SN 2003bg have higher 
velocities, with the velocity at $\sim$ 40 days being $\sim$ 13500 km sec$^{-1}$, 
SN 2011dh at $\sim$ 12000 km sec$^{-1}$ is intermediate, and SN 1993J has the
lowest at $\sim$ 9500 km sec$^{-1}$. A similar trend is seen in the evolution 
of the H$\beta$ line velocity also. This may be related to the mass of the 
hydrogen envelope at the time of explosion, and the explosion energy.  
\cite{iwamoto97} have shown that for type IIb supernovae, the minimum velocity 
of hydrogen depends mainly on the mass of the hydrogen envelope. This indicates
that the mass of hydrogen envelope in SN 2011dh is between SN 1993J 
(0.2 to 0.9 M$_\odot$) and SN 2003bg (0.05 M$_\odot$). The estimate of hydrogen envelope 
mass of 0.1 M$_\odot$ in SN 2011dh by \cite{bresten12} agrees with this.  

\subsubsection{Oxygen mass  and [Ca\,II]/[O\,I] ratio}

The nebular spectra of stripped envelope core-collapse supernovae can be used 
to estimate mass of oxygen which is an indicator of the mass of the progenitor.
\cite{uomoto86} has shown that in the high density limit ($N_{e} \ge 10^{6}$ 
cm$^{-3}$), the minimum mass of  oxygen required to produce the observed 
[O\,I] emission can be esimated using the relation 
\begin{equation}
M_{O} = 10^{8} \times D^{2} \times F([O\,I]) \times exp^{(2.28/T_{4})},
\end{equation} 
 
where, $M_{\rm{O}}$ is the mass of neutral oxygen in M$_{\odot}$, $D$ is the 
distance in Mpc, F([O\,I]) is the flux of the [O\,I] 6300, 6364~\AA\ line in 
ergs sec$^{-1}$ and $T_{4}$ is the temperature of the oxygen emitting region in 
units of 10$^{4}$ K. Ideally, estimation  of $T_{4}$ should be  done using the 
flux ratio of [O\,I] 5577~\AA\ to [O\,I] 6300-6364~\AA\ lines. However, the
presence of a weak [O\,I] 5577~\AA\ line in the nebular spectra shows that the 
temperature in the oxygen emitting region is low and the limit of flux ratio 
[O\,I] 5577/6300-6364 can be assumed to be $\le$ 0.1.  In this limit, the 
emitting region will be either at high density ($N_{e} \ge 10^{6}$ cm$^{-3}$) 
and low temperature ($T_{4} \le 0.4$) or at low density ($N_{e} \le 10^{6}$ 
cm$^{-3}$) and high temperature ($T_{4} = 1.0$) \citep{maeda07}. For the few 
well studied type Ib supernovae, during the nebular phase, the oxygen emitting 
region is usually found to be of high density and low temperature 
(\citealt{leibundgut91}, \citealt{elmhamdi04}, \citealt{schlegel89}). Using 
the observed flux of 1.03 $\times$ 10$^{-13}$ erg sec$^{-1}$ cm$^{-2}$ of 
[O\,I]6300, 6364~\AA\ line on 2012 February 26  and  assuming that 
$T_{4} = 0.4$~K holds good for SN 2011dh, the mass of oxygen is estimated as 
0.22 M$_\odot$. There is a hint of the presence of [O\,I] 7774~\AA\ line in
the late phase spectra of SN 2011dh, this line arises mainly due to the 
recombination of ionized oxygen \citep{begelman86}. Hence, it is appropriate 
to consider that some oxygen lies in an ionized form and the estimated mass of 
oxygen 0.22 M$_\odot$  is a lower limit of total mass of oxygen ejected during 
the explosion.   

The estimate of mass of oxygen obtained, using a similar methodology, for some 
type Ib/c  supernovae indicates that it varies from $\sim$ 0.3 - 1.35 M$_\odot$ 
\citep{elmhamdi04}. For SN 2007Y, \cite{stritzinger09} 
have shown that in their model  an oxygen mass of 0.2 M$_\odot$ reproduces 
the observed late time spectrum.  Specifially, for type IIb events, 
\cite{houck96}  arrived at an oxygen mass of $\sim$ 0.5 M$_\odot$ for SN 1993J
by modelling the late time spectra, whereas modelling of early and late 
time spectra of SN 2003bg led \cite{mazzali09} to an oxygen mass estimate of 
1.3 M$_\odot$. The mass of oxygen derived by modelling the late time spectra 
of SN 2001ig is $\sim$ 0.8 M$_\odot$ \citep{silverman09}. The oxygen mass
estimated for SN 2011dh is smaller as compared to the well studied type IIb 
events SN 1993J, SN 2003bg, and is similar to the mass ejected in the explosion 
of SN 2007Y. 

Oxygen is ejected mostly from the oxygen layer formed during the hydrostatic 
burning phase, and so, the mass of oxygen ejected is very sensitive to the main 
sequence mass of the progenitor.  \cite{thielemann96} have made 
explosive nucleosynthesis  calculation and predicted major nucleosynthesis  
yields for the progenitor mass 
of 13 - 25 M$_\odot$. They have shown that mass of  ejected oxygen is 0.22, 
0.43, 1.48 and 3.0 M$_\odot$ for progenitor mass of  13, 15, 20 and 25 
M$_\odot$, respectively. \cite{nomoto06} have calculated 
nucleosynthesis yields of core-collapse supernovae and hypernovae models for 
13 - 40 M$_\odot$ progenitor 
stars for various explosion energies and progenitor metallicity. For 
core-collapse supernovae of a progenitor mass 13 M$\odot$, with the explosion 
energy of  $1\times 10^{51}$ ergs and metallicity Z = 0, 0.001, 0.004 and 
0.02, the mass of oxygen ejected in the explosion is 0.45, 0.50, 0.39 and 
0.22 M$\odot$, respectively.  For similar  explosion energy and metallicity 
range, the mass of oxygen for a progenitor star of 15 M$\odot$ is  0.77, 0.29, 0.29 
and 0.16 M$\odot$. The abundance analysis of H\,II regions in M51 by \cite{bresolin04} 
has revealed that the O/H  abundance is below the solar value for most of the H\,II 
regions studied.  H\,II regions \#53, \#54 and \#55 of  \cite{bresolin04}  are close 
to the location where SN 2011dh occurred. The measured 12 + log(O/H) values for 
these regions varies between 8.49 to 8.66 (considering the solar value as 8.69). 
It shows that the metallicity  of the nearby H\,II regions is close to few tenths 
of that of solar. Assuming that the metallicity at the supernova location is similar 
to the nearby H\,II regions, the mass of the ejected oxygen estimated using the oxygen 
flux in the late time spectra,  indicates  the progenitor  of SN 2011dh to be a 
low-mass star of $\sim$ 13 - 15  M$\odot$. The progenitor of a type IIb supernova  
can be either a single massive star  that had gone to Wolf-Rayet phase 
after losing most of its hydrogen rich envelope before explosion, or it can be 
a less massive star in a binary system,  
being stripped off its hydrogen envelope during interaction with the binary 
component. The inferred lower  mass of  the progenitor suggests  that the progenitor 
of SN 2011dh was likely a  member of a binary system like the progenitor of 
SN 1993J (\citealt{podsiadlowski93}, \citealt{maund09}) and SN 2008ax (\citealt{roming09}, \citealt{taubenberger11}).      

\cite{fransson89}  have theoretically calculated the ratio of flux of 
[Ca\,II] 7291-7324/[O\,I] 6300-6364 and shown that it weakly depends on the density 
and temperature of the emitting region, and is expected to remain relatively constant at late epoch. The flux ratio 
[Ca\,II] 7291-7324/[O\,I] 6300-6364 also serves as a good diagnostic of main 
sequence mass of the progenitor star, as  mass of oxygen ejected in the 
explosion depends on the  progenitor mass 
and  mass of  Ca synthesized during the explosion is not sensitive to the 
main sequence mass of the progenitor \citep{nomoto06}. Hence, a small value of 
the flux ratio is expected for massive progenitors.  
\cite{elmhamdi04} have investigated  the evolution of observed [Ca\,II]7291-7324/[O\,I]6300-6364 for some core 
collapse supernovae  during sufficiently late phases and  shown that it remains fairly stable. The smaller  
value of [Ca\,II]/[O\,I] ratio for type Ib/c supernova  was explained as due to the absence of hydrogen 
rich Ca\,II emitting region in them.  For SN 2011dh, the  [Ca\,II]7291-7324/[O\,I]6300-6364 flux ratio at 
$\sim$ 340 and $\sim$ 360 days after explosion is found to $\sim$ 0.7, indicating
that the progenitor star of SN 2011dh was a low mass star.  The flux ratio for 
SN 1993J, SN 2008ax and SN  2007Y at similar epochs is $\sim$ 0.5, 0.9 and 1.0, respectively.  

\cite{bresten12}  have computed hydrodynamical models based on evolutionary 
progenitors and shown that the 
early light curve of SN 2011dh can be reproduced with a large progenitor star with radius $\sim$ 200 R$\odot$. 
In their modelling,  based on  the bolometric light curve and measured photospheric expansion velocity, mass 
of the ejecta,  explosion energy and mass of $^{56}$Ni was constrained to be $\sim$ 2 M$\odot$, 
$6-10 \times 10^{50}$ erg and   0.06 M$\odot$, respectively. An explosion of a   progenitor star with  
a He core of 3 to 4 M$\odot$ and a thin hydrogen rich envelope of $\sim$ 0.1 M$\odot$ can lead to 
SN 2011dh like event.  The progenitor star  appears to be in a binary system with the main sequence mass 
 between $12-15$ M$\odot$. \cite{benvenuto13} have shown that evolution of a  binary system with 
16 M$\odot$ + 10 M$\odot$, and an initial period of 125 days, appears compatible with the pre-SN observations 
of SN 2011dh, and predicted that the yellow super giant star detected in the pre-SN images could be the 
potential progenitor candidate.  Our direct estimates of  mass of $^{56}$Ni using the bolometric light curve, 
and property of the progenitor inferred using the  mass of oxygen ejected in the explosion, the [Ca\,II]/[O\,I] 
ratio in the nebular phase and  mass of hydrogen envelope are in very good 
agreement with the binary models.     

\section{Conclusions}

In this paper, we present optical photometry in $UBVRI$ bands and medium 
resolution spectroscopy of the type IIb supernova SN 2011dh in nearby galaxy 
M51, during 3 days to one year after the explosion. SN 2011dh reached a peak 
absolute $B$ magnitude of $M_{B} = -16.378\pm0.18$ with a rise time of 
$19.6\pm0.5$ days. With an absolute $V$ peak magnitude of 
$M_{V} = -17.123\pm0.18$, SN 2011dh is $\sim$ 0.3 magnitude fainter than the mean 
absolute magnitude
of a sample of type IIb supernovae. The initial decline in magnitude within 15 
days from the date of maximum ($\Delta m_{15}$) in the $B$ and $V$ bands is 
found to be similar to those of SN 1993J and SN 2008ax. In the late phase, 
between days 170 to 360, a steepening is seen in the $B$-band light curve, while
the $R$ and $I$ bands show a flattening.  This indicates the possibility
of dust formation. The colour evolution shows that the 
$(B-V)$ and $(V-R)$ colours of SN 2011dh are always redder as compared to those
of other stripped envelope core-collapse supernovae, which could be because of  
lower temperature of the supernova.  Simple analytic fits to the bolometric 
light curve of indicate that $\sim$ 0.09 M$_\odot$ of 
$^{56}$Ni was synthesized in the explosion.

The early spectral evolution of SN 2011dh shows prominent P-Cygni absorption 
due to Balmer lines of hydrogen, the H$\alpha$ absorption line is seen in the 
spectrum till $\sim$ 60 days after explosion. The He\,I lines start appearing 
in the spectrum much before the maximum in $B$-band. Synthetic spectra have 
been computed using the SYN++ code and the main features in the spectra have 
been identified.  The early photospheric temperature estimated at $\sim 7500$~K
is cooler than the estimates for other SNe IIb, indicating a more conducive
condition for early dust formation. The emergence of forbidden lines due to 
[O\,I] 6300, 6363~\AA\ and [Ca\,II] 7291, 7324~\AA\ in the spectrum of 
$\sim 100$ days after explosion indicates the onset of the nebular phase. The 
[O\,I] 6300-6364 line shows a double peaked profile. A box shaped emission due 
to H$\alpha$ is seen in the red wing of the [O\,I] line, which could possibly 
arise because of the shock-wave interaction of CSM. The ejected mass of oxygen 
($\sim 0.22$~M$_\odot$), estimated using the [O\,I] 6300, 6363~\AA\ line flux, 
together with an intermediate value of the [Ca\,II] 7291-7324/[O\,I] 6300-6364 
 flux ratio suggests a less massive progenitor in a binary system. 

\section{Acknowledgement}
We would like to thank the referee  for a careful reading of the 
manuscript and constructive suggestions, which helped in improving the 
manuscript.
We thank all the observers of the 2-m HCT (operated by Indian Institute of As
rophysics), who kindly provided part of their observing time for observations
of the supernova. We thank  K. Nomoto and K. Maeda  for
useful discussion. 
This work has made use of the NASA/IPAC Astrophysics Data System and the NASA/IPAC Extragalactic
Database (NED) which is operated by Jet Propulsion Laboratory, California Institute of Technology,
under contract with the National Aeronautics and Space Administration.

\end{document}